\documentclass[preprint,12pt,compress]{elsarticle}

\usepackage{amssymb}
\usepackage{amsmath}
\usepackage{bm}
\usepackage{subfigure} 
\usepackage{epstopdf}
\usepackage{url}
\usepackage{float}
\usepackage[colorlinks, linkcolor=blue, citecolor=blue]{hyperref}
\journal{Measurement}

\begin{document}

\begin{frontmatter}



\title{Depth estimation of a monoharmonic source using a vertical linear array at fixed distance} 


\author[1]{Yangjin Xu}
\author[1]{Wei Gao}
\author[1]{Xiaolei Li\corref{corresponding}}
\ead{lxl@ouc.edu.cn}
\author[1]{Qinghang Zeng}

\cortext[corresponding]{Corresponding author. School of Marine Technology, Ocean University of China, Qingdao, 266100, China}

\affiliation[1]{
	organization={School of Marine Technology, Ocean University of China},
	city={Qingdao},
	postcode={266100},
	country={China}
}

\begin{abstract}
Estimating the depth of a monoharmonic sound source at a fixed range using a vertical linear array (VLA) is challenging in the absence of seabed environmental parameters, and relevant research remains scarce. The orthogonality constrained modal search based depth estimation (OCMS-D) method is proposed in this paper, which enables the estimation of the depth of a monoharmonic source at a fixed range using a VLA under unknown seabed parameters. Using the sparsity of propagating normal modes and the orthogonality of mode depth functions, OCMS‑D estimates the normal mode parameters under a fixed source‑array distance at first. The estimated normal mode parameters are then used to estimate the source depth. To ensure the precision of the source depth estimation, the method utilizes information on both the amplitude distribution and the sign (positive/negative) patterns of the estimated mode depth functions at the inferred source  depth. Numerical simulations evaluate the performance of OCMS-D under different conditions. The effectiveness of OCMS-D is also verified by the Yellow Sea experiment and the SWellEx-96 experiment. In the Yellow Sea experiment, the depth estimation absolute errors by OCMS-D with a 4-second time window are less than 2.4 m. And the depth estimation absolute errors in the SWellEx-96 experiment with a 10-second time window are less than 5.4 m for the shallow source and less than 10.8 m for the deep source.
\end{abstract}




\end{frontmatter}


\section{Introduction}
\label{sec1}
Acoustic methods for localizing a source, particularly an underwater source is a topic of great interest. As an important part of source localization, the depth estimation is of great significance to identify underwater sources. When the direct and reflected signals from a sound source are sufficiently strong, methods based on the interference structure \cite{ref7,ref8} can accurately estimate the source depth. However, such methods are only applicable to broadband sources, and require the range between the source and the hydrophones to be relatively short. For sound sources such as distant vessels traveling through shallow-water waveguides, line spectra are generally less vulnerable to noise masking than continuous spectra, owing to their higher spectral level and the potential for signal to noise ratio (SNR) improvement via temporal integration. When only the line spectra are observable, the distant vessels can be regarded as harmonic sources, and the methods based on interference structure are not applicable. Conventional matched field processing (CMFP) \cite{ref9,ref10,ref11,ref12} is theoretically capable of estimating the depth of a harmonic source. However, the accuracy of CMFP-based source depth estimation method heavily depends on the precision of ocean environmental modeling. When environmental model errors are present, CMFP is unable to accurately estimate the source depth. In reality, obtaining accurate environmental parameters, particularly seabed parameters, poses significant challenges, and it often leads to severe environmental mismatch problem in practical applications of CMFP. To address the environmental mismatch problem, T. C. Yang proposed a data-based matched mode processing (DMMP) method \cite{ref13,ref14}. Unlike the CMPF method, DMMP synthesizes a virtual horizontal array by utilizing a moving source to extract normal mode parameters, enabling the estimation of monoharmonic source depth with only a single hydrophone. Since the normal mode parameters are extracted directly from measured data, DMMP effectively avoids the environmental model mismatch problem inherent in CMPF. However, as DMMP primarily exploits acoustic field information along the horizontal range and is not designed for real-time depth estimation, it cannot perform target depth estimation using a vertical linear array (VLA) under the condition of a fixed source range. For a sound source at a fixed distance, one need to explore other depth estimation methods. In recent years, deep learning has been widely applied in underwater acoustics, offering new approaches for underwater source depth estimation \cite{ref20,ref21,ref22}. However, deep learning-based methods for source depth estimation often require substantial training data and exhibit limited generalization capability \cite{ref22.1}. When the test environment significantly differs from the training environment, the performance of source depth estimation deteriorates rapidly. Furthermore, the interpretability of deep learning-based methods is a difficult problem requires more research. Recently, a seabed independent depth estimation (SIDE) method has been proposed by the authors using the modal Doppler effect \cite{ref23}, which does not require prior knowledge of source frequency or velocity and can estimate the depth of harmonic sources under unknown seabed parameters. Nevertheless, the SIDE method requires the source to move a considerable distance to estimate depth, and it also demands that the signal radiated by the source contains at least two distinct frequencies. In reality, targets seldom radiate two stable line-spectrum signals over extended durations. If target depth estimation can be achieved using a single-frequency signal under a relatively fixed source range, it would not only circumvent the inherent problems associated with the SIDE method but also facilitate underwater target identification and real-time tracking.  However, there exists limited literature on this subject to the best of our knowledge, rendering it an open challenge. 

In this paper, a method called orthogonality constrained modal search based depth estimation (OCMS-D) is proposed for monoharmonic source depth estimation. It integrates the normal mode parameter estimation method orthogonality constrained modal search (OCMS) \cite{ref24}  and the SIDE method developed in our prior work, thereby achieving depth estimation of a monoharmonic source using a VLA under conditions of unknown seabed parameters and a fixed source range. Simulations are conducted to analyze the performance of OCMS-D under different conditions. The effectiveness of OCMS-D is also validated using experimental data from the Yellow Sea experiment and the SWellEx-96 experiment  \cite{ref25}.

The structure of this paper is organized as follows. Section~\ref{sec2} provides the theory of the OCMS-D. The performance of the OCMS-D under different conditions is evaluated by simulations in Section~\ref{sec3}. Validation results using experimental data from the Yellow Sea experiment and the SWellEx-96 experiment are demonstrated in Section~\ref{sec4}. Finally, a conclusion is given in Section~\ref{sec5}.

\section{Theory of OCMS-D}
\label{sec2}

To estimate the depth of a monoharmonic source, it is necessary to estimate the mode depth functions of normal modes and the amplitude distribution of each mode excited by the source from the received acoustic field data. Therefore, this section first provides a brief review of the OCMS method, detailing how it estimates both the mode depth functions and the mode amplitude distribution using a VLA. Building on this foundation, it then describes how to utilize the estimated mode depth functions and the normal mode amplitude distribution to achieve source depth estimation.

\subsection{Mode estimation using a VLA}
\label{subsec2.1}

This subsection is an brief review of the OCMS method, which combines the sparsity of received normal modes and the orthogonality of mode depth functions to estimate the mode parameters under conditions of fixed source-VLA distance and unknown seabed parameters. Should any readers wish to gain an in-depth understanding of the OCMS, they are encouraged to refer to Ref. \cite{ref24}.

In a range independent ocean waveguide with depth $H$, assume the sound speed profile (SSP) $c\left( z \right)$ of the water column is known. When the normal mode wavenumber $\xi_m$ is given, the corresponding mode depth function $\psi(z,\xi_m)$ can be obtained by the following difference equation \cite{ref27}
\begin{equation}
\begin{cases}
	\psi_m (z + h, \xi_m) + \psi_m (z - h, \xi_m) = 2 \cos \left( h \sqrt{k^2(z) - \xi_m^2} \right) \psi_m (z, \xi_m)\\
    \psi_m(0,\xi_m)=0\\
    \sum_{l=0}^L|\psi_m(lh,\xi_m)|^2h=1
\end{cases},
	\label{eq4}
\end{equation}
where $z\in[0,H]$, $Lh\in(H-h,H]$, $h\ll1$, $k(z)=\omega/c(z)$, and $m$ can be determined by the number of nodes of $\psi_m(z,\xi_m)$. Although $M$ mode depth functions require $M$ normal mode wavenumbers to determine in general, under the assumption that the mode depth functions in water column are orthogonal to each other, all mode depth functions in water column are determined when any normal mode wavenumber is given. Here, the orthogonality of the mode depth functions in water column can be expressed by the following equation:
\begin{equation}
\label{eq4a}
\begin{array}{c}
\sum_{l=0}^{L} \psi_{m}\left(z_{l}, \xi_{m}\right) \psi_{n}^{*}\left(z_{l}, \xi_{n}\right) h=\delta_{m n},
\end{array}
\end{equation}
where $z_l=lh$, ``$\ast$'' represents complex conjugate, and $\delta_{mn}$ is the Kronecker delta function.  Give a normal mode wavenumber $\xi_m$, a set of mode depth functions, 
\[
\begin{array}{l}
\Psi(z,\xi_m ):= \{\psi_1 (z,\xi_1 ),\psi_2 (z,\xi_2 ),\cdots,\psi_m (z,\xi_m ),\cdots,\psi_M (z,\xi_M ) \} ,
\end{array}
\]
can be obtained by combing Eqs (\ref{eq4}) and (\ref{eq4a}), where $M$ is the number of considered normal modes. The procedure for obtaining $\Psi(z,\xi_m)$ from $\xi_m$ has been detailed in \cite{ref24} and is not repeated here for brevity. 

Now, consider a VLA containing $N$ elements with depths ${{\bf{z}}_r} = \left[ {{z_1},{z_2}, \ldots, {z_N}} \right]$ is deployed in the water. The pressure received by the VLA from a distant monoharmonic source can be expressed as a finite linear combination of elements from the set of mode depth functions, i.e.

\begin{equation}
     {\bf{p}} = {\bf{\Phi }}\left( {{k_1}} \right){\bf{a}}\left( {{k_1}} \right)
     \label{eq1}
\end{equation}
where
\begin{equation*}
	{\bf{p}} = {\left[ {p\left( {{z_1}} \right),p\left( {{z_2}} \right), \ldots p\left( {{z_N}} \right)} \right]^{\rm{T}}},
\end{equation*}
\begin{equation*}
	{\bf{\Phi }}\left( {{k_1}} \right) = \left[ {{\bm{\phi} _1}\left( {{k_1}} \right),{\bm{\phi} _2}\left( {{k_2}} \right), \ldots ,{\bm{\phi} _M}\left( {{k_M}} \right)} \right],
\end{equation*}
\begin{equation*}
	{\bm{\phi} _m}\left( {{k_m}} \right) = \left[ {{\phi _m}\left( {{z_1},{k_m}} \right),{\phi _m}\left( {{z_2},{k_m}} \right), \ldots ,{\phi _m}\left( {{z_N},{k_m}} \right)} \right]^{\rm{T}} \ {\rm{with}}\ m = 1,2, \ldots ,M,
\end{equation*}
\begin{equation*}
	{\bf{a}}\left( {{k_1}} \right) = {\left[ {{a_1}\left( {{k_1}} \right),{a_2}\left( {{k_2}} \right), \ldots ,{a_M}\left( {{k_M}} \right)} \right]^{\rm{T}}},
\end{equation*}
$k_m$ is the $m{\rm{th}}$ modal wavenumber and ${\phi _m}\left( {{{{z}}},{k_m}} \right)$ is the mode depth function at depth $z$ corresponding to $k_m$. Both $k_m$ and $\phi_m(z,k_m)$ can be calculated by the  KRAKEN program \cite{ref30}.  Noting that the “$k_1$” in ${\bf{\Phi }}\left( {{k_1}} \right)$  and ${\bf{a}}\left( {{k_1}} \right)$  serves as a label, 
\begin{equation*}
	{\bf{\Phi }}\left( {{k_1}} \right) = {\bf{\Phi }}\left( {{k_2}} \right) =  \ldots  = {\bf{\Phi }}\left( {{k_M}} \right),
\end{equation*}
\begin{equation*}
	{\bf{a}}\left( {{k_1}} \right) = {\bf{a}}\left( {{k_2}} \right) =  \cdots  = {\bf{a}}\left( {{k_M}} \right),
\end{equation*}
and
\begin{equation}
	{a_m}\left( {{k_m}} \right) = \frac{i}{4}S\left( \omega  \right){\phi _m}\left( {{z_s},{k_m}} \right)H_0^{\left( 1 \right)}\left( {{k_m}r} \right)
	\label{eq2}
\end{equation}
being the mode amplitude of the $m{\rm{th}}$ order mode. In Eq.~\eqref{eq2}, $S\left( \omega  \right)$ is the source spectrum, $\omega  = 2\pi f$ is the angular frequency of the radiated signal, $z_s$ is the source depth, $r$ is the horizontal distance between the VLA and the source. Then the OCMS method shows that the normal mode parameters can be estimated by the following equation:

\begin{equation}
	\hat{k}_m = \mathop {\arg \min }\limits_{\xi_m}   \left\| \mathbf{a}(\xi_m)\right\|_1, 
	\label{eq7}
\end{equation}
subject to
\begin{equation}
    \left\| \mathbf{p} - {\bf{\Psi}}(\xi_m)\mathbf{a}(\xi_m)\right\|_2 < \epsilon_n,\nonumber
\end{equation}
where $\hat{k}_m$ and $\bm{\Psi}(\hat{k}_m)$ denotes the estimations of $k_m$ and $\bm{\Phi}({k}_m)$ respectively, 
 \[
 \mathbf{\Psi}(\xi_m)=[\bm{\psi}_1 (\xi_1 ),\bm{\psi}_2 (\xi_2 ),\cdots,\bm{\psi}(\xi_m), \cdots,\bm{\psi}_M (\xi_M )],
 \]
\[
\bm{\psi}_m (\xi_m )=[\psi_m (z_1,k_m ),\psi_m (z_2,k_m ),\cdots,\psi_m (z_N,k_m )]^\mathrm{T},
 \]
$\mathbf{a}(\hat{k}_m)$ presents the estimation result of $\mathbf{a}({k}_m)$,  and $\epsilon_n > 0$ is determined by the SNR. This paper utilizes the noise energy in the VLA to take the value of $\epsilon_n$.  It should be noted that both $\bm{\Phi}(k_m)$ and $\bm{\Psi}(\xi_m)$ represent the matrices of mode depth functions at the depths of the VLA. Since the VLA may not be densely populated throughout the water column, the column vectors of  $\bm{\Phi}(k_m)$ and $\bm{\Psi}(\xi_m)$ are not necessarily orthogonal to each other.  Eq.~\eqref{eq7} can be solved using the CVX toolbox \cite{ref28}. This method enables simultaneous estimation of the modal wavenumbers, mode depth functions, and corresponding complex mode amplitudes for a fixed monoharmonic source, which paves the way for estimating the depth of an underwater monoharmonic source at unknown seabed parameters. 

\subsection{Depth estimation with the mode sign}
\label{subsec2.2}

When ${\bf{\Psi}}(\hat{k}_m)$ and  ${\bf{a}}(\hat{k}_m)$ are obtained, source depth can be estimated by constructing a depth ambiguity function \cite{ref11}.  

\begin{equation}
	D(z) = \left| \sum_{m=1}^{M} \psi_m (z,\hat{k}_m) | a_m (\hat{k}_m) | \delta_m \right|^2,
	\label{eq8}
\end{equation}
where $\delta_m = \pm 1$ is used to compensate for the signs of the mode amplitudes. Substituting Eq.~(\ref{eq2}) into the depth ambiguity function yields \cite{ref11}  
\begin{equation}
	D(z) \sim \left| \sum_{m=1}^{M} \psi_m (z,\hat{k}_m) \psi_m (z_s,\hat{k}_m) \right|^2 \sim \delta(z - z_s).
	\label{eq9}
\end{equation}

Obviously, the depth ambiguity function exhibits a maximum peak when $z = z_s$, where the peak position corresponds to the source depth. However, the results of Eq.~\eqref{eq9} are based on the premise that the signs of the mode amplitudes are correctly compensated, else it may result in significant errors in the depth estimation. Therefore, some studies assign values to specific scenarios, for instance, all   $\delta_m$ can be set to positive for sources near the sea surface \cite{ref11}, or alternatively, the absolute values of both the mode amplitudes and the mode depth functions are taken without considering mode signs \cite{ref23}. The former is not applicable to sources at other depths, while the latter ignoring mode signs will face the issue of high sidelobes in the depth ambiguity function. To address the challenge of estimating mode signs, we have proposed a method called depth-sign search (DSS), which can efficiently estimate the mode signs. The basic principles of the depth estimation method based on DSS will be briefly described next. 

According to Eq.~\eqref{eq2}, the mode signs depend on the signs of the mode depth functions at the source depth, which is generally positive or negative except at a few nodes. Since the source depth is unknown, the mode signs are initially undetermined. Nevertheless, given that the mode depth functions have already been estimated, the signs of different modes at the same depth can be combined. Using the property that the correctly signs compensated depth ambiguity function approximates a Dirichlet kernel function  \cite{ref29}, a cost function is constructed. And a search is then performed over the depth dimension to find the mode signs that most closely resemble the depth ambiguity function and the Dirichlet kernel function, which is taken as the estimate of the mode signs at the source depth.

The $m$th mode sign at depth $z_q$ can be obtained by  
\begin{equation}
	\delta_m(q) = \text{sign} \left[ \psi(z_q, \hat{k}_m) \right],
	\label{eq10}
\end{equation}
where $q = 1, \cdots, Q$, $Q$ is the number of combinations that distinguish mode signs along depth, and $\text{sign} \left[ \bullet \right]$ denotes taking the sign of the variable. The cost function measuring similarity of the depth ambiguity function and the Dirichlet kernel function is defined using the Kullback-Leibler (KL) divergence:  
\begin{equation}
	KL(q) = \int_0^H D(z, q) \ln \frac{D(z, q)}{Ds(z, q)} dz,
	\label{eq11}
\end{equation}
where  
\begin{equation}
	D(z, q) = \frac{\left| \sum_{m=1}^M \psi_m(z, \hat{k}_m) |a_m(\hat{k}_m)| \delta_m(q) \right|^2}{\int_0^H\left| \sum_{m=1}^M \psi_m(z,\hat{k}_m) |a_m(\hat{k}_m)| \delta_m(q) \right|^2dz}
	\label{eq12}
\end{equation}
represents the normalized depth ambiguity function compensated by the $q$th mode signs combination,  
\begin{equation}
	Ds(z, q) = \frac{\big|\frac{\sin \left[ (M+1)\pi(z - z_q)/H \right]}{\sin \left[ \pi(z - z_q)/(2H) \right]}\big|^2 + \epsilon}{\int_0^H\big\{\big|\frac{\sin \left[ (M+1)\pi(z - z_q)/H \right]}{\sin \left[ \pi(z - z_q)/(2H) \right]}\big|^2 + \epsilon\big\} dz}
	\label{eq13}
\end{equation}
represents the normalized theoretical depth ambiguity function under the $q$th mode signs combination \cite{ref23}, and $\epsilon$ is a small constant (which can typically be taken as $10^{-6}$) added to avoid $Ds(z,q) = 0$. The correct mode signs combination is then defined as

\begin{equation}
	{\bf{\delta }}({q_0}) = \left[ {{\delta _1}({q_0}),{\delta _2}({q_0}), \ldots ,{\delta _M}({q_0})} \right],
	\label{eq14}
\end{equation}
where $q_0 = \arg \min_q KL(q)$. Furthermore, the estimated source depth is

\begin{equation}
	z_s = \mathop {\arg \max }\limits_z D(z,q_0)
	\label{eq15}
\end{equation}

It should be noted that the estimated mode signs combination also corresponds to a depth $z_{q_0}$. Since a single signs combination may correspond to multiple depths (a scenario more likely to occur when the number of modes is small), $z_{q_0}$ is typically not used as the estimated source depth. Instead, the result of Eq.~\eqref{eq15} is used as the estimated source depth. A more detailed discussion on this issue can be found in \cite{ref23}.

Notably, as previously mentioned, OCMS-D can estimate the depth of a monoharmonic source without the seabed parameters and source motion. In addition, VLA is not required to cover the entire water column. Subsequent numerical simulations will provide a relevant analysis.

\section{Numerical simulations}
\label{sec3}

\begin{figure}[htbp]
	\centering
	\includegraphics[height=6cm]{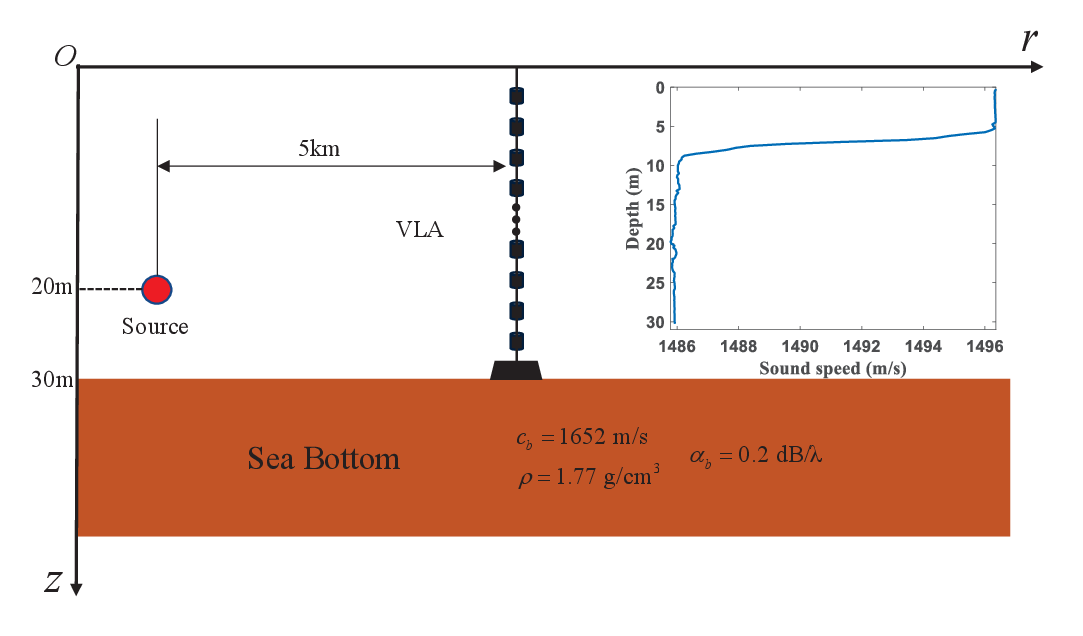}
	\caption{The parameters of a range-independent waveguide, along with a VLA and a monoharmonic source within the waveguide.}\label{fig1}
\end{figure}

This section evaluates the performance of OCMS-D through numerical simulations. The SSP used in the simulation was collected in the Yellow Sea during spring, as shown in Fig.~\ref{fig1}. It contains a thermocline which linearly decreases from 1496 m/s at 8 m to 1485 m/s at 10 m, and remains nearly constant down to the sea bottom at 31 m. The seabed sound speed is 1652 m/s, the density is 1.77 $\text{g/cm}^3$, and the attenuation coefficient is 0.2 dB/$\lambda$ , where  $\lambda$ represents wavelength.

\subsection{Results of mode estimation and depth estimation}
\label{subsec3.1}

As shown in Fig.~\ref{fig1}, a VLA comprising $N=30$ elements with an element spacing of 1 m is deployed in the water column. Unless otherwise specified, the first element is fixed at a depth of 1 m below the surface in the following simulations. A monoharmonic source at depth $z_s=20$~m radiates a monoharmonic signal at 596~Hz. And the distance $r$  between the source and the VLA is 5~km. Under these conditions, the signal received by the VLA predominantly contains 10 modes. In data processing, all parameters except the SSP are assumed unknown, and $\xi_m\in[\xi_{\rm{min}},\xi_{\rm{max}}]$ with
\begin{equation}
    \xi_{\max} = 2\pi f/[\min(c(z))] = 2.5217~\rm{m^{-1}},\nonumber
\end{equation}
\begin{equation}
    \xi_{\min} = 2\pi f/[\max(c(z))+300~\rm{m/s}] = 2.0851~\rm{m^{-1}},\nonumber
\end{equation}
to ensure that $\bm{\Psi}(\xi_m)$ in Eq.~\eqref{eq7} encompasses a sufficient number of normal modes. The pressure received by the VLA is simulated using the KRAKEN program \cite{ref30}, with 30 dB noise added. And the pressure with noise can be obtained by 
\begin{equation}
	{{\mathbf{p}_{\rm{nosie}}}={\bf{p}}+{\bf{n}}},
	\label{eq16.1}
\end{equation}
where ${\mathbf{p}_{\rm{nosie}}} \in {\mathbb{C}^{N \times 1}}$ and ${\bf{n}} = [{n_1},{n_2}, \ldots ,{n_N}]^\mathrm{T}$ represent the pressure vector and the noise vector, respectively. Here, $\mathbb{C}$ represents the complex number domain, and ${{n_l}}$, where $l=1,2,\cdots,N$, are mutually independent and each follows a complex Gaussian distribution with a mean of $0$ and a variance of $\sigma^2$.  Then, the SNR is defined as follows.
\begin{equation}
	{\rm{SNR = 20log_{10}}}\frac{{{{\left\| {\bf{p}} \right\|}_2}}}{{{N\sigma}}},
	\label{eq16.2}
\end{equation}

\begin{figure}[htbp]
	\centering  
	\setlength{\abovecaptionskip}{0.cm}
	\subfigure[] 
	{
		\begin{minipage}[b]{.45\linewidth} 
			\centering
			\includegraphics[height=5cm]{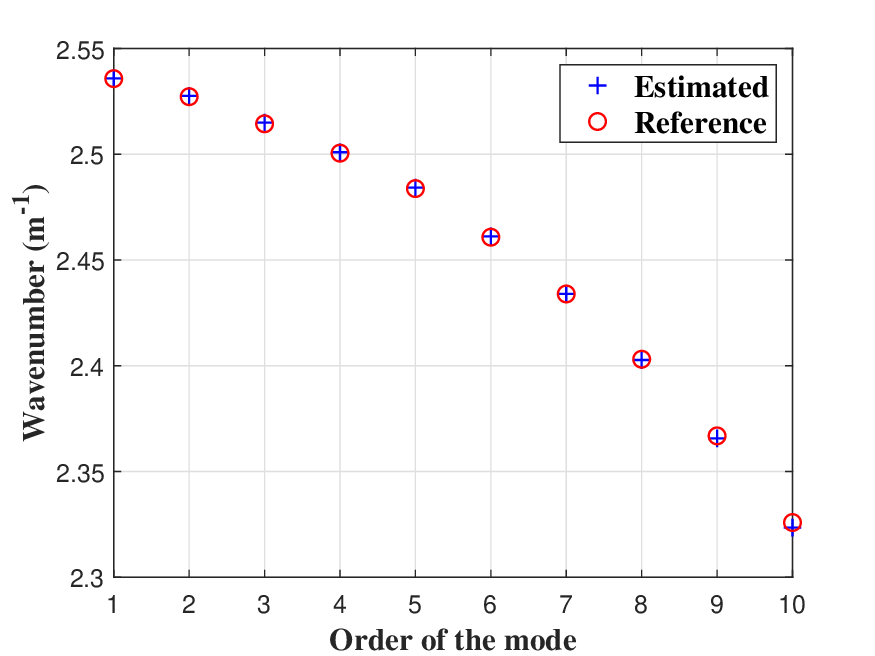}
			\vskip -5pt
			\label{fig2a}
		\end{minipage}}
	\subfigure[]
	{
		\begin{minipage}[b]{.45\linewidth}
			\centering
			\includegraphics[height=5cm]{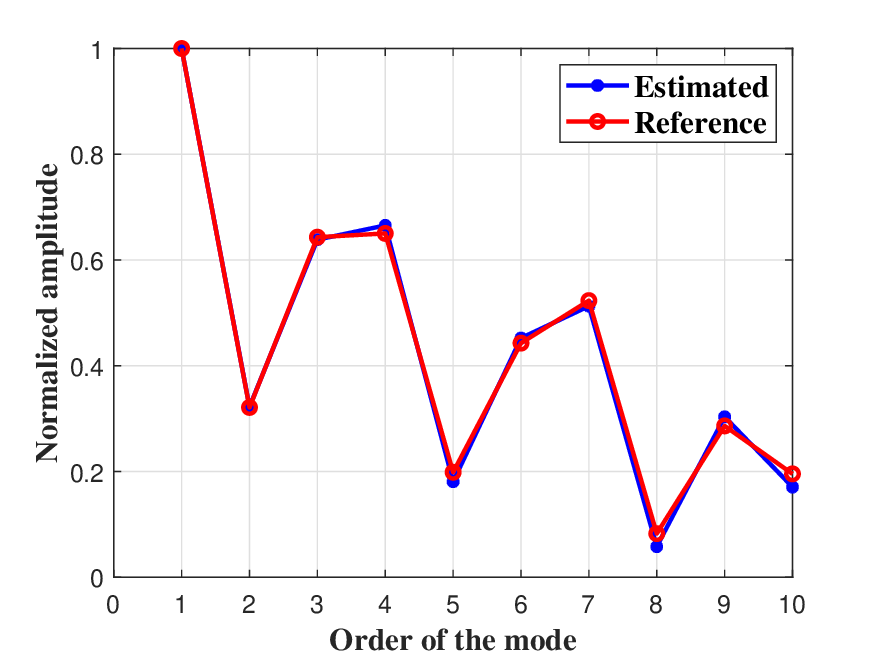}
			\vskip -5pt
			\label{fig2b}
		\end{minipage}
	}
	
	\subfigure[]
	{
		\begin{minipage}[b]{1\linewidth}
			\centering
			\includegraphics[height=5cm]{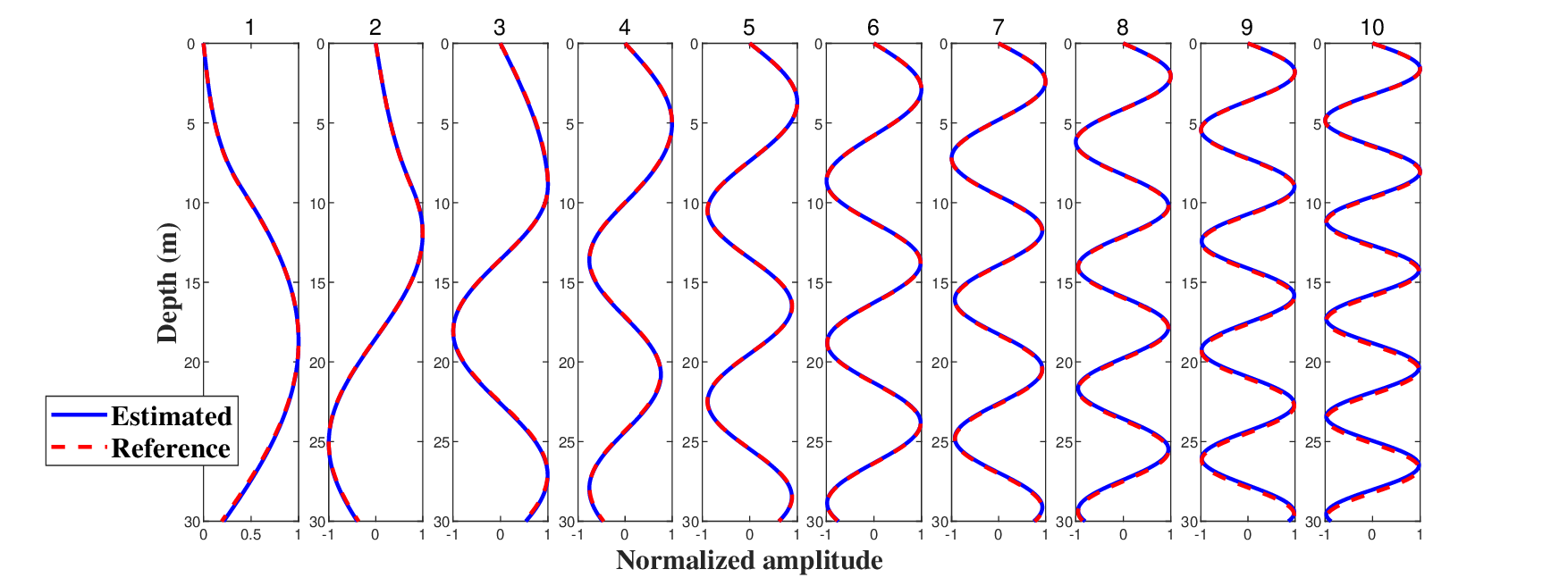}
			\vskip -5pt
			\label{fig2c}
		\end{minipage}
	}
	\caption{Comparison of estimated horizontal wavenumbers, mode amplitudes (magnitude), and mode depth functions with Kraken-calculate reference values, respectively. Red markers indicate reference values, while blue markers represent estimated values.}
	\label{fig2}
\end{figure}

Using the OCMS, the estimated modes are obtained, as shown in Fig.~\ref{fig2a}-~\ref{fig2c}. Fig.~\ref{fig2a}-~\ref{fig2c} compares the estimated horizontal wavenumbers (blue crosses), mode amplitudes (blue line with dots), mode depth functions (blue line) with the Kraken-calculated reference values (red labels), respectively. Fig.~\ref{fig2a}-~\ref{fig2c} indicated that the estimated results align well with the reference values. From Fig.~\ref{fig2} (a), it can be observed that the estimation error of the normal mode wavenumber increases with the order of the mode. This occurs because the OCMS method assumes orthogonality of the mode depth functions within the water column. However, higher-order normal modes are more prone to energy penetration into the seabed, which weakens the orthogonality assumption for the corresponding mode depth functions, thereby leading to larger estimation errors in the wavenumbers of these modes.

\begin{figure}[htbp]
	\centering
	\setlength{\abovecaptionskip}{0.cm}
	\includegraphics[height=6cm]{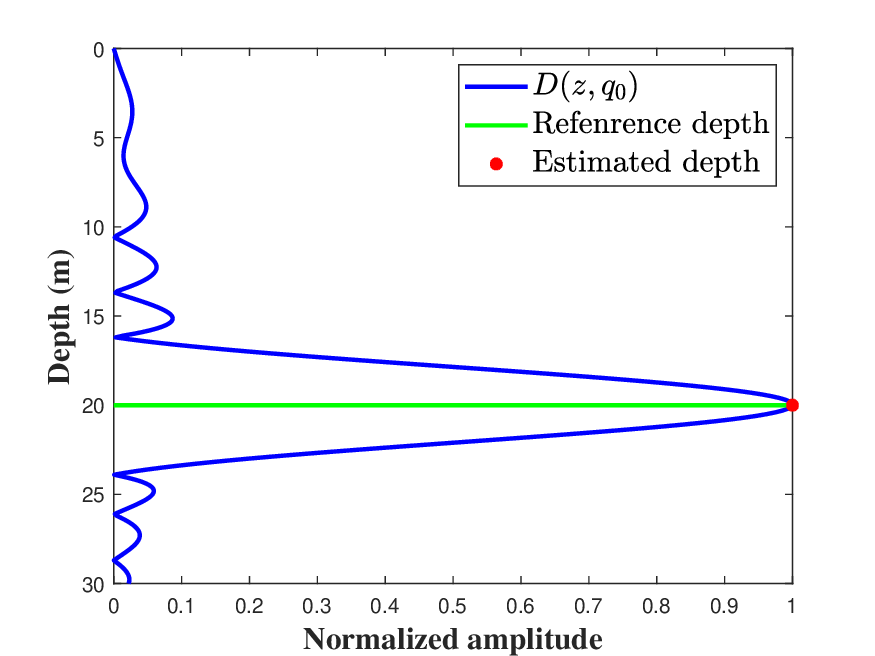}
	\caption{Depth ambiguity function  $D(z,q_0)$ (blue line) obtained using OCMS-D and the reference depth (green line). The result of depth estimation ${\hat z_s} = 20$m.}\label{fig3}
\end{figure}

The signs of the estimated mode depth functions are then determined using Eq.~\eqref{eq10}. In this paper, the sign is sampled at an interval of 0.1 m, i.e., $Q = H/0.1$ . These values are substituted into Eq.~\eqref{eq11} and Eq.~\eqref{eq12} to estimate the mode signs. Subsequently, the source depth estimation result is derived using Eq.~\eqref{eq15}. Fig.~\ref{fig3} shows the depth ambiguity function compensated by the mode signs combination corresponding to the minimum value of the loss function (blue line). The result of depth estimation ${\hat z_s} = 20$ m, which aligns exactly with the reference depth (green line).

\subsection{Different SNR}
\label{subsec3.2}

\begin{figure}[htbp]
	\centering   
	\setlength{\abovecaptionskip}{0.cm}
	\subfigure[]
    {
		\begin{minipage}[b]{.45\linewidth} 
			\centering
			\includegraphics[height=5cm]{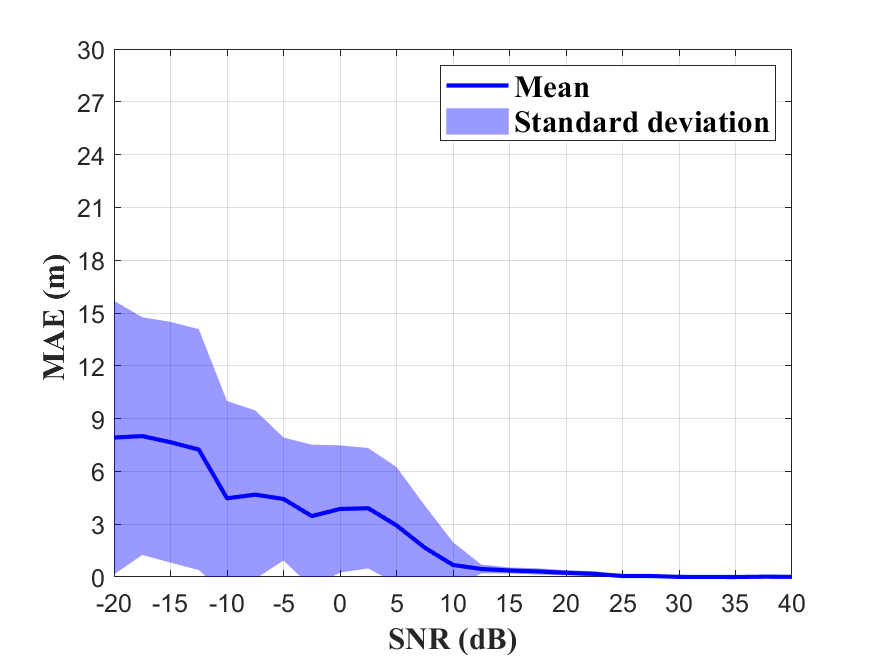}
			\vskip -5pt
			\label{fig4a}
		\end{minipage}
	}
    \subfigure[] 
	{
		\begin{minipage}[b]{.45\linewidth} 
			\centering
			\includegraphics[height=5cm]{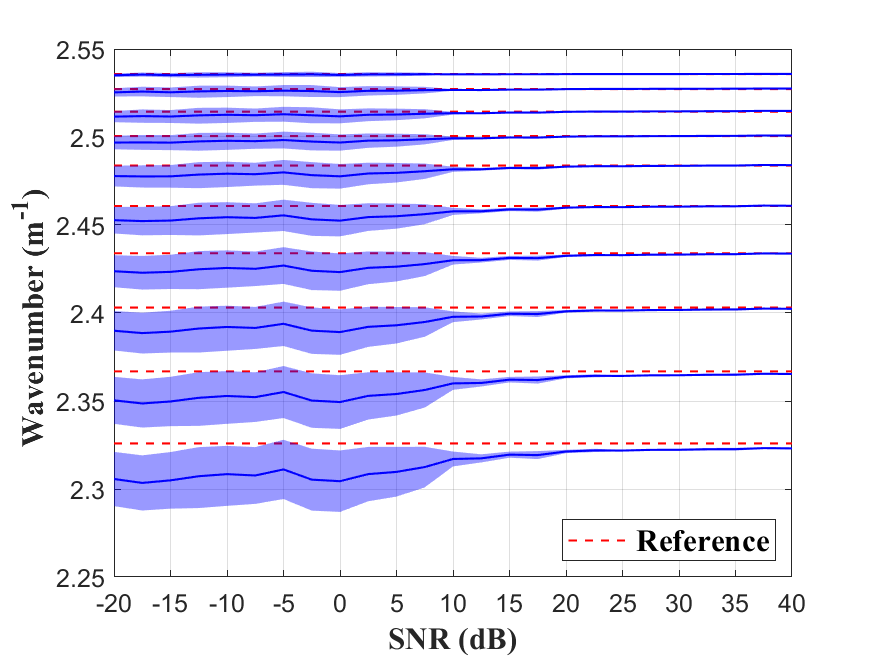}
			\vskip -5pt
			\label{fig4b}
		\end{minipage}
	}
	\subfigure[]
	{
		\begin{minipage}[b]{.45\linewidth}
			\centering
			\includegraphics[height=5cm]{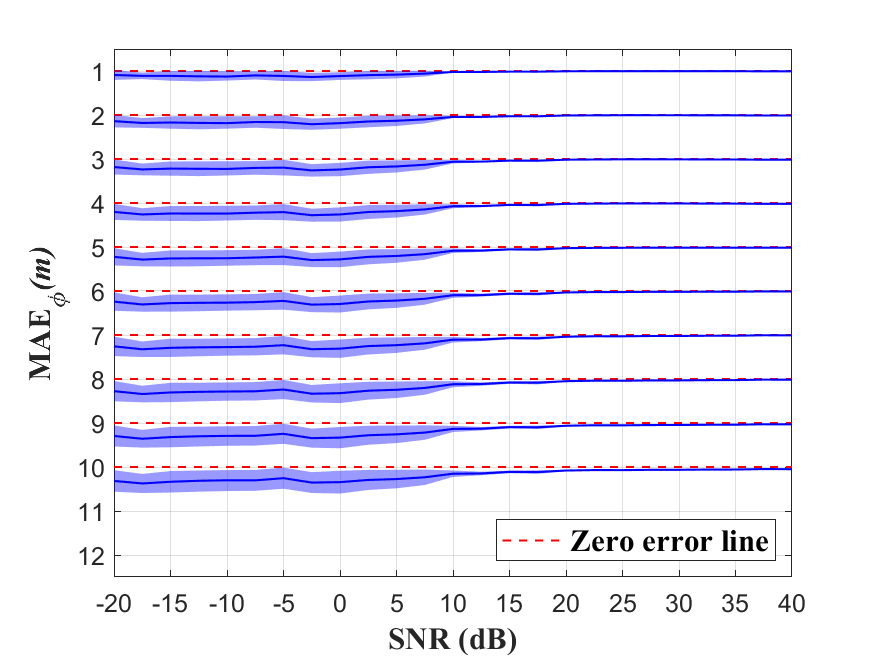}
			\vskip -5pt
			\label{fig4c}
		\end{minipage}
	}
    \subfigure[]
	{
		\begin{minipage}[b]{.45\linewidth}
			\centering
			\includegraphics[height=5cm]{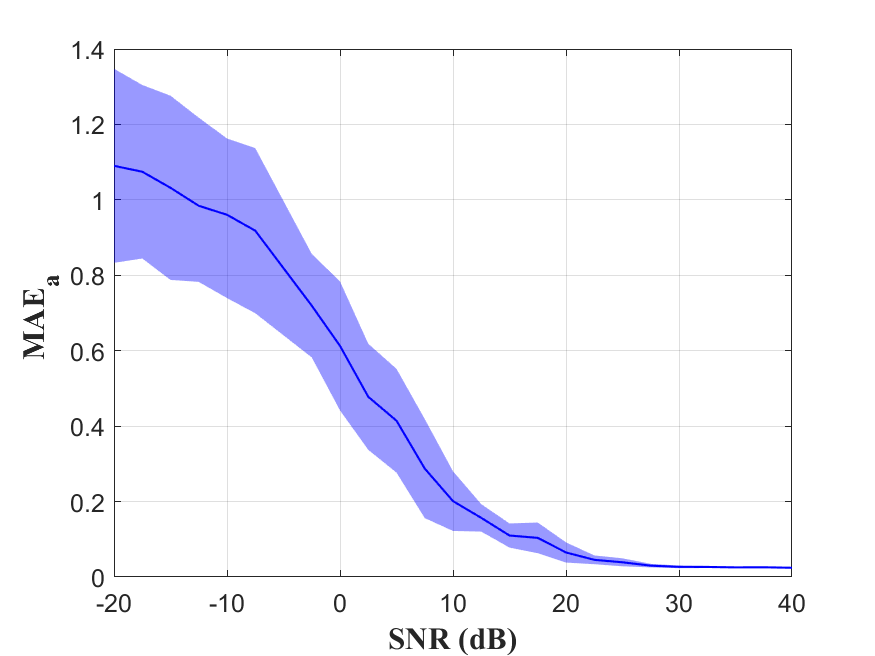}
			\vskip -5pt
			\label{fig4d}
		\end{minipage}
	}
	\caption{Estimation results of source depth and modal parameters, along with their standard deviations, obtained by OCMS-D under different SNRs. (a) Estimation results of source depth. (b) Estimation results of wavenumbers. The red dashed lines indicates the Kraken-calculate reference value. (c) Estimation results of the mode depth functions. The red dashed lines indicates the zero error lines. (d) Estimation results of modal amplitudes. The blue lines and blue shaded regions in figures represent the mean and standard deviation intervals of the estimated results, respectively. }
	\label{fig4}
\end{figure}

This subsection analyzes the performance of OCMS-D under different SNR conditions. The SNR varies from $-20$ dB to 40 dB with increments of 2.5 dB. For each SNR condition, $J=50$ independent Monte Carlo trials are conducted to evaluate the mean absolute error (MAE) and standard deviation of the depth estimation results. The absolute error of the $j\rm{th}$ Monte Carlo trial and the MAE are defined by 
\begin{equation}
    \begin{cases}
	\text{AE}_j &= \left| {{{z_s} - {{\hat z}_{sj}}}} 
    \right| \\
   {\text{MAE}} &= \frac{1}{J}\sum\limits_{j = 1}^J {{\text{AE}_j}}
   \end{cases}
   \label{eq17},
\end{equation}
where $\hat{z}_{sj}$ represents estimated value of $z_s$ at the $j\rm{th}$ Monte Carlo trial. The VLA configuration is kept consistent with that in the previous simulation. Results are shown in Fig.~\ref{fig4}.  Fig.~\ref{fig4a} presents the MAE of source depth under different SNRs. The blue lines indicate the mean estimated values, and the blue shaded regions represent the standard deviation intervals. This applies to the other figures in Fig.~\ref{fig4} as well. Since the estimated wavenumber error is small (maximum magnitude of 1e-2), the estimation results under different SNRs are directly presented in Fig.~\ref{fig4b}.  Fig.~\ref{fig4b} also shows the wavenumbers calculated by the KRAKEN program as reference. Fig.~\ref{fig4c} presents the estimated errors for each mode depth function estimated by OCMS-D under different SNRs. To more clearly illustrate the impact of different orders of mode depth function errors estimated by OCMS-D as SNR varies, the errors are added to their respective order of the mode, i.e.,
\begin{equation}
{\rm{MAE}_\phi }\left( m \right) = \frac{1}{{JH}}\sum\limits_{j = 1}^J {\int_0^H {\left| {{\psi _m^{j}}\left( {z,{{\hat k}_m}} \right) - \phi_m \left( {z,{k_m}} \right)} \right|dz} }  + m, 
\end{equation}
where ${\psi _m^j}\left( {z,{{\hat k}_m}} \right)$ denotes the estimated value of mode depth function $\phi_m \left( {z,{k_m}} \right)$ at the $j\rm{th}$ Monte Carlo trial. The red dashed lines mark the zero‑error line in Fig.~\ref{fig4c} . Similarly, Fig.~\ref{fig4d} shows the estimated errors of each modal amplitude vector, which defined by 
\begin{equation}
{\rm{MA}}{{\rm{E}}_a} = \frac{1}{J}\sum\limits_{j = 1}^J \|\mathbf{a}_j(\hat{k}_m)-\mathbf{a}(k_m)\|_2,
\end{equation}
where $\mathbf{a}_{j}\left( {{{\hat k}_m}} \right)$ denotes the estimated value of  ${\mathbf{a}}\left( {{k_m}} \right)$ at the $j\rm{th}$ Monte Carlo trial, and $\|\mathbf{a}(k_m)\|_2=1$. As observed in Figs. \ref{fig4b}-\ref{fig4d},  the estimation errors of normal mode wavenumbers,  mode depth functions, and ${\mathbf{a}}\left( {{k_m}} \right)$  all increase as the SNR decreases, which consequently leads to a growth in the target depth estimation error with declining SNR. However, it can be noted that the estimation errors of the normal mode parameters rise only gradually as SNR decreases. This slow growth explains why the depth estimation error remains relatively small even under low SNR conditions.

\subsection{Different VLA parameters}
\label{subsec3.3}

\begin{figure}[htbp]
	\centering   
	\setlength{\abovecaptionskip}{0.cm}
	\subfigure[] 
	{
		\begin{minipage}[b]{.45\linewidth} 
			\centering
			\includegraphics[height=5cm]{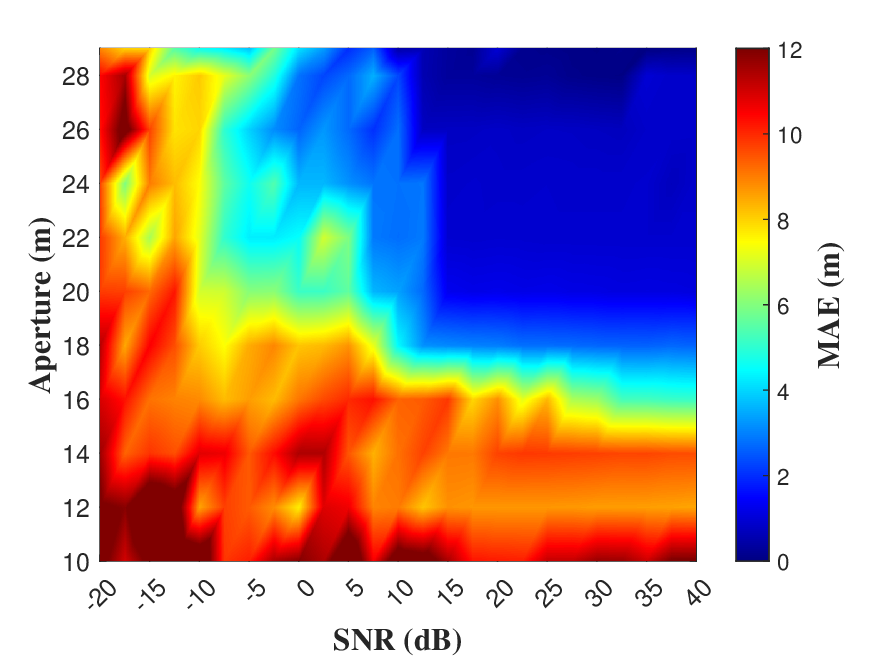}
			\vskip -5pt
			\label{fig5a}
		\end{minipage}
	}
	\subfigure[]
	{
		\begin{minipage}[b]{.45\linewidth}
			\centering
			\includegraphics[height=5cm]{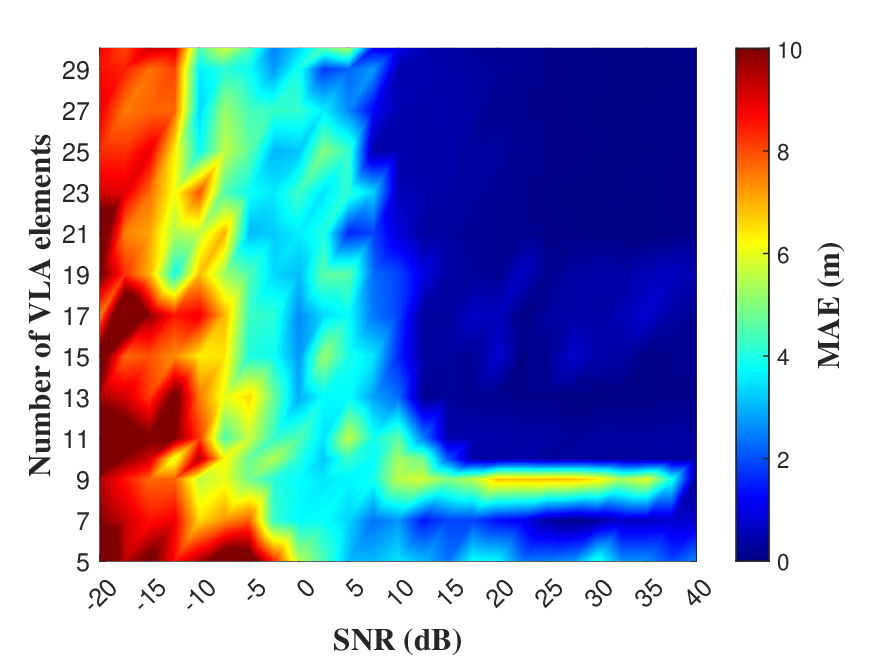}
			\vskip -5pt
			\label{fig5b}
		\end{minipage}
	}
	\subfigure[] 
	{
		\begin{minipage}[b]{.45\linewidth} 
			\centering
			\includegraphics[height=5cm]{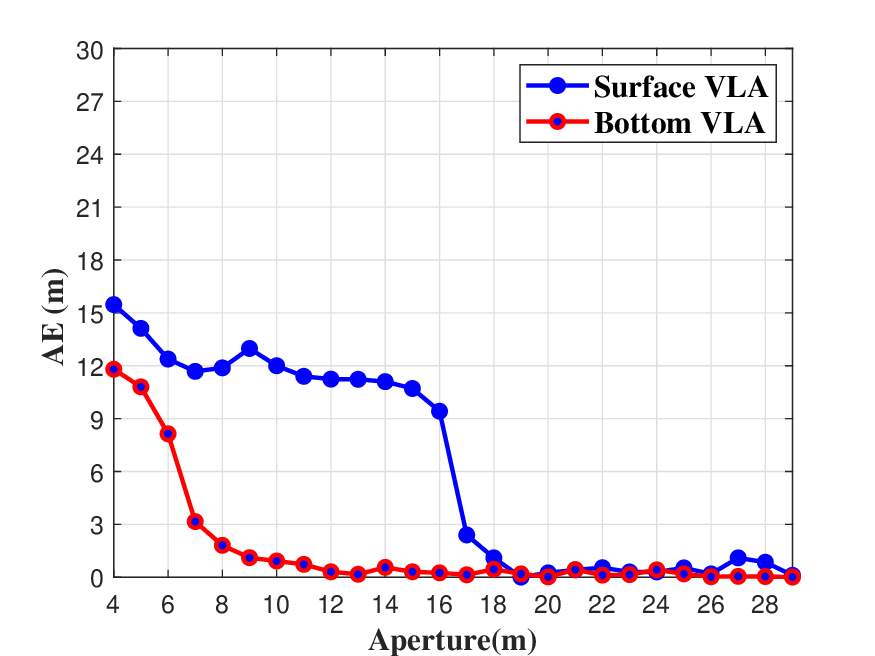}
			\vskip -5pt
			\label{fig5c}
		\end{minipage}
	}
	\subfigure[]
	{
		\begin{minipage}[b]{.45\linewidth}
			\centering
			\includegraphics[height=5cm]{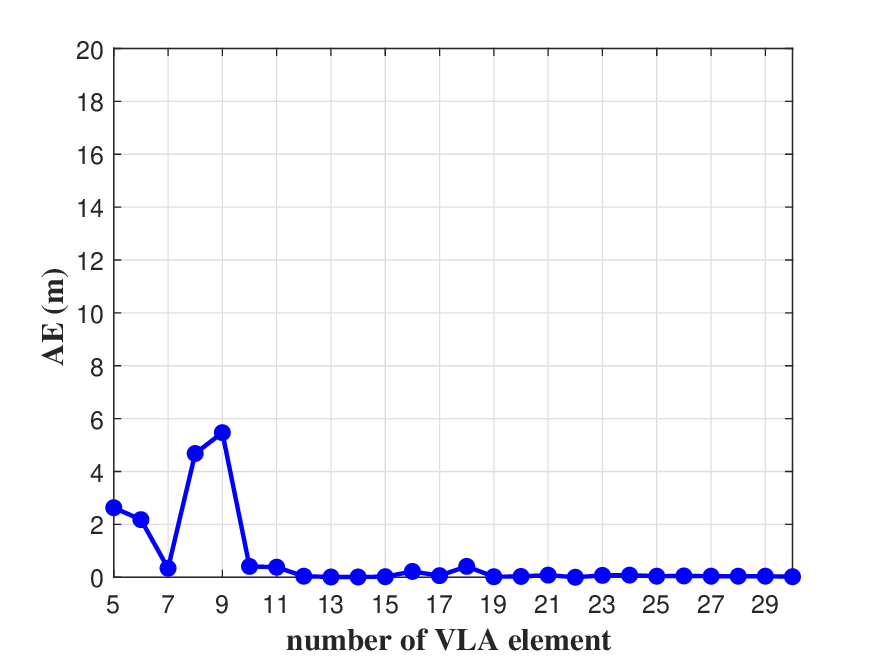}
			\vskip -5pt
			\label{fig5d}
		\end{minipage}
	}
	\caption{The AE of depth estimation results under different array parameters and SNR. (a) and (b) are the MAE for the SNR-Aperture plane and SNR-Number plane, respectively. The blue lines in (c) and (d) are the results for SNR=30 dB in (a) and (b), respectively. The red lines in (c) is the results for a bottom VLA when the SNR = 30 dB.}
	\label{fig5}
\end{figure}

It is well known that the parameters of a VLA significantly influence array-based depth estimation methods. Therefore, this subsection evaluates the performance of OCMS-D under different VLA parameters and SNRs by varying array aperture, number of array elements and SNRs. The SNR parameter settings are the same as in Subsection 3.2. And the array aperture is varied from 4 m to 29 m in increments of 1 m while maintaining a fixed number of 30 elements. It should be reiterated that the first element is deployed at a depth of 1 m below the surface. Fig.~\ref{fig5a} presents the MAE of the results for a source at a depth of 20 m under different array apertures and SNRs. The blue dotted line in Fig.~\ref{fig5c} corresponds to the result for SNR = 30 dB in Fig.~\ref{fig5a}. It can be observed that OCMS‑D achieves better performance in this simulation environment when the array aperture exceeds 18~m and the SNR is higher than 7.5~dB. Furthermore, with the array aperture fixed at 29 m, the number of elements was increased from 5 to 30 in steps of 1, and the resulting AE is shown in Fig.~\ref{fig5b}. Similarly, Fig.~\ref{fig5d} displays the result for SNR = 30 dB extracted from Fig.~\ref{fig5b}. From Fig.~\ref{fig5b} and Fig.~\ref{fig5d}, it is evident that OCMS‑D performs better in this simulation setup when the number of elements is greater than 9 and the SNR is higher than 7.5~dB. These simulation results also confirm that OCMS‑D does not require the vertical line array to cover the entire water column.

Moreover, this subsection also includes simulations comparing the depth estimation performance of OCMS-D when the VLA is deployed on the sea bottom versus at the sea surface. The simulation environment parameters were consistent with those specified earlier, with a SNR set to 30 dB. And the deepest element of the VLA was fixed at the bottom, the array aperture was varied from 29 m down to 4 m in steps of 1 m. The simulation results are shown as the red line in Fig.~\ref{fig5c}. Compare the results of the surface VLA and the bottom VLA, one can find that the bottom VLA can achieve favorable depth estimation performance with a relatively short effective array aperture.

\subsection{Different number of modes}
\label{subsec3.4}
\begin{figure}[htbp]
	\centering   
	\setlength{\abovecaptionskip}{0.cm}
	\subfigure[] 
	{
		\begin{minipage}[b]{.45\linewidth} 
			\centering
			\includegraphics[height=5cm]{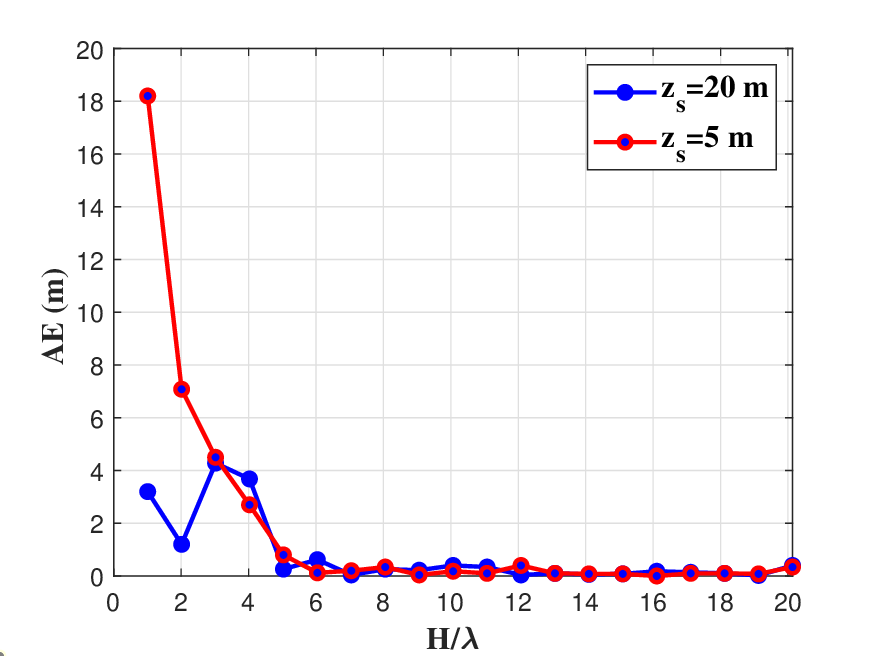}
			\vskip -5pt
			\label{fig5_5a}
		\end{minipage}
	}
	\subfigure[]
	{
		\begin{minipage}[b]{.45\linewidth}
			\centering
			\includegraphics[height=5cm]{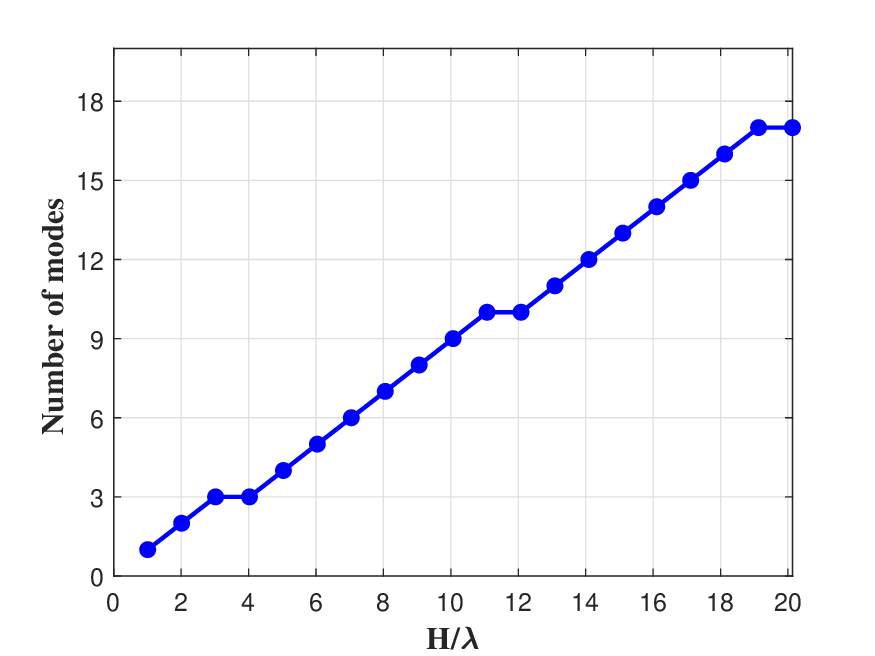}
			\vskip -5pt
			\label{fig5_5b}
		\end{minipage}
	}
	\caption{(a) The AE of depth estimation results under different source frequency. (b) The number of modes excited by the source under different frequency.  }
	\label{fig5_5}
\end{figure}

The number of propagating normal modes excited by the source in the waveguide, as well as the accuracy of their estimation, is critical for depth estimation performance. The number of propagating normal modes is jointly determined by the water depth and the source frequency. Higher frequency or greater water depth leads to more propagating normal modes. To evaluate the performance of OCMS-D under varying number of normal modes, this subsection examines the behavior of OCMS-D across different number of modes by adjusting the source frequency within the oceanic environment described in Section 3.1. In the simulation, the source frequency was varied from 50 Hz to 1000 Hz in steps of 50 Hz, with the SNR set to 30 dB and the VLA parameters consistent with those in Subsection 3.1. The simulation results are presented in the Fig.~\ref{fig5_5}. Fig.~\ref{fig5_5a} represents the AE of depth estimation results under different source frequencies.
 And Fig.~\ref{fig5_5b} represents the number of modes excited by the source under different frequency. It can be observed that under this simulation environment, the depth estimation performance of OCMS-D becomes unstable at $H/\lambda$ below 5 (corresponding to 250 Hz and 4 modes). 
This is primarily because lower frequencies excite fewer propagating normal modes in the waveguide, making it difficult to achieve reliable depth estimation through the depth ambiguity function. For instance, when $H/\lambda = 1$, only the first-order mode is excited. Under this condition, the depth estimation result will always correspond to the peak position of the first-order mode depth function. As a result, the estimation error varies with the source depth. This is illustrated in Fig.~\ref{fig5_5}, where the depth estimation errors for the deep and shallow sources differ significantly when $H/\lambda = 1$.

\section{Experiment data analysis}
\label{sec4}

The effectiveness of OCMS-D will be validated using the Yellow Sea experiment and the SWellEx-96 experiment in this section. 

\subsection{Yellow Sea experiment}
\label{subsec4.1}

\begin{figure}[htbp]
	\centering  
	\setlength{\abovecaptionskip}{0.cm}
	\subfigure[] 
	{
		\begin{minipage}[b]{.45\linewidth} 
			\centering
			\includegraphics[height=4.8cm]{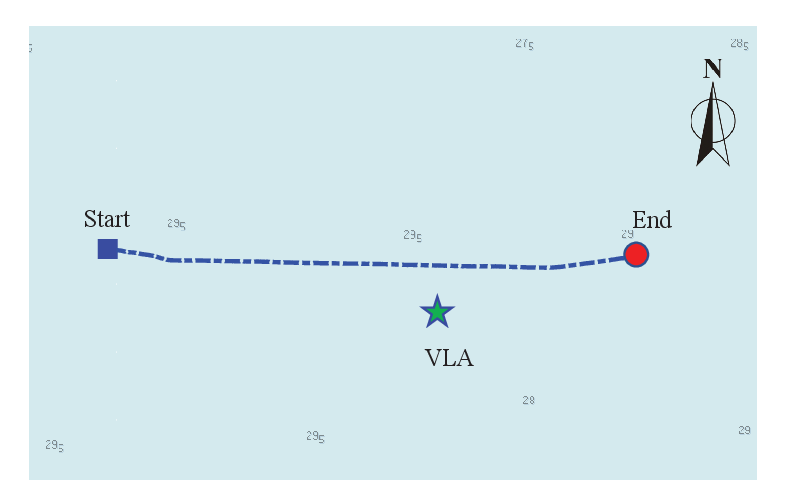}
			\vskip -3pt
			\label{fig6a}
		\end{minipage}
	}
	\subfigure[]
	{
		\begin{minipage}[b]{.45\linewidth}
			\centering
			\includegraphics[height=5cm]{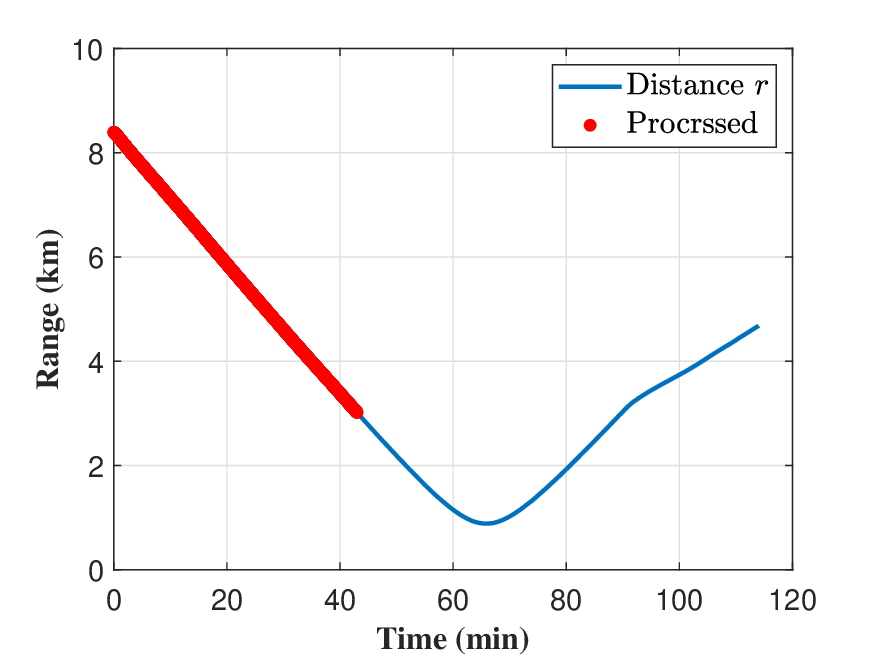}
			\vskip -5pt
			\label{fig6b}
		\end{minipage}
	}
	
	\subfigure[]
	{
		\begin{minipage}[b]{.45\linewidth} 
			\centering
			\includegraphics[height=5cm]{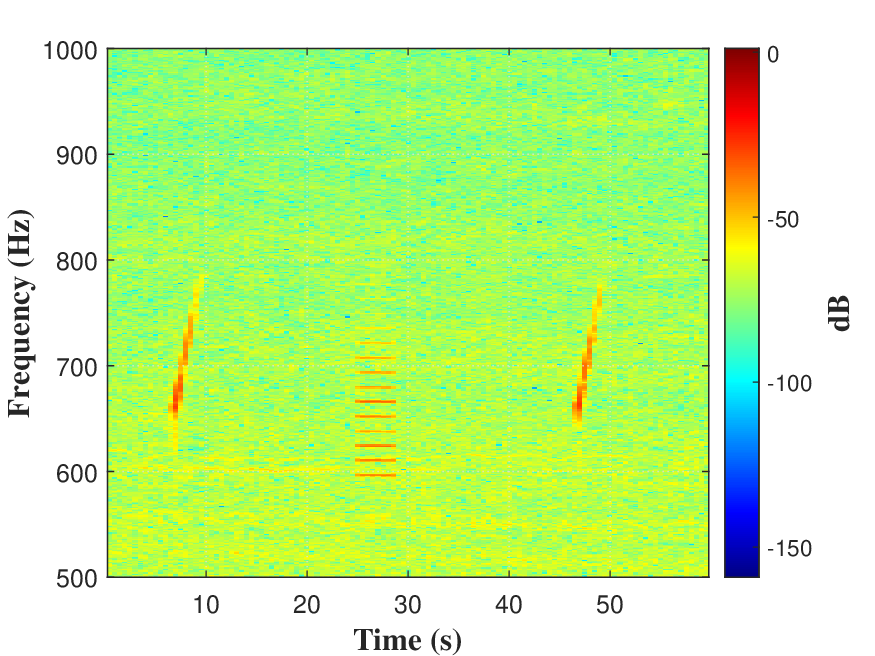}
			\vskip -5pt
			\label{fig6c}
		\end{minipage}
	}
	\subfigure[]
	{
		\begin{minipage}[b]{.45\linewidth}
			\centering
			\includegraphics[height=5cm]{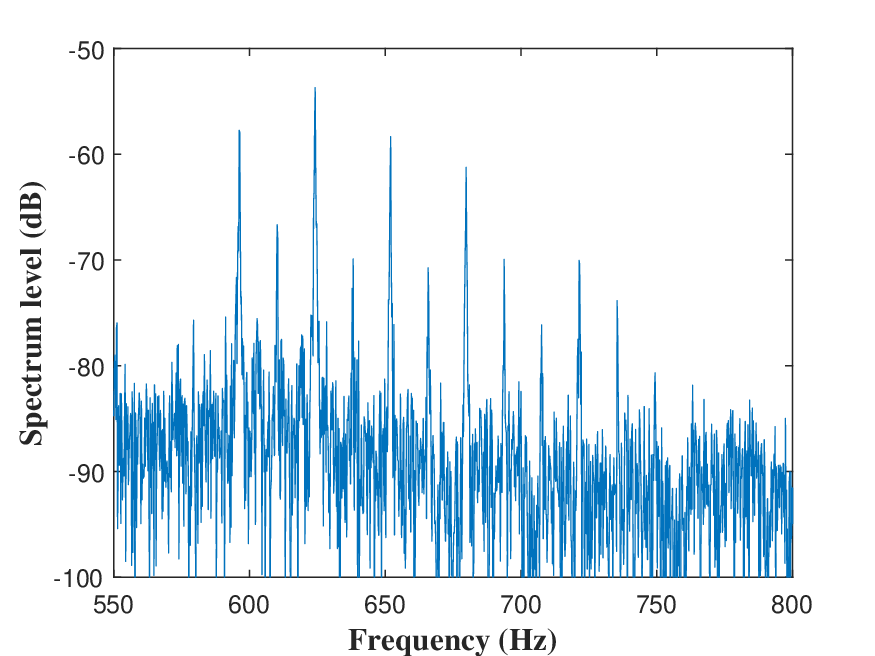}
			\vskip -5pt
			\label{fig6d}
		\end{minipage}
	}
	\caption{(a) The trajectory of the source ship and the VLA position during the Yellow Sea experiment. (b) The range between the source ship and the VLA. The red dots indicate the time period processed in this subsection. (c) Time-frequency diagram of the signal emitted by the sound source. (d) Spectrum of the CW signal received by the hydrophone closest to the water surface in the VLA (spectrum analysis duration is 4 seconds).}
	\label{fig6}
\end{figure}

The trajectory of the source ship and the VLA position during the Yellow Sea experiment are shown in Fig.~\ref{fig6a}, with a water depth of approximately 31 m. A sound source at a depth of 19 m underwater was towed by the ship, which traveled from west to east at a speed of about 4 knots and passed through the VLA. The distance between the source ship and the VLA, as recorded by GPS, is shown in Fig.~\ref{fig6b}, and the SSP measured during the experiment is shown in Fig.~\ref{fig1}. The source alternately emitted linear frequency-modulated (LFM) signals and continuous wave (CW) pulses at a depth of approximately 19 m until the experiment finished. Both signal types had a duration of 4 s, separated by 16 s intervals, as depicted in Fig.~\ref{fig6c}. The frequency of CW signal ranged from 596 Hz to 792 Hz, incrementing by 14 Hz, as shown in Fig.~\ref{fig6d}. A 10-element VLA with 2 m element spacing was deployed south of the source trajectory, and the first element was located at a depth of 3 m below the surface. It should be noted that the simulation environment parameters in Section~\ref{sec2} primarily reference those of the Yellow Sea experiment. This subsection will validate OCMS-D using the 596 Hz frequency data from the CW signal, as the simulation results in Section~\ref{sec2} indicate that the VLA can receive 10 modes at this frequency.

\begin{figure}[htbp]
	\centering   
	\setlength{\abovecaptionskip}{0.cm}
    \subfigure[] 
	{
		\begin{minipage}[b]{.45\linewidth} 
			\centering
			\includegraphics[height=5cm]{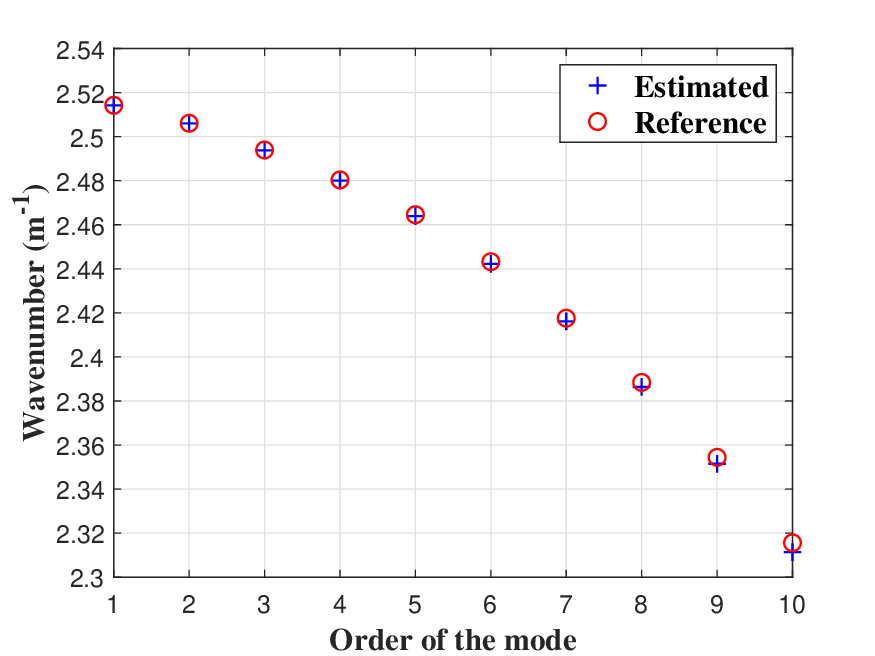}
			\vskip -5pt
			\label{fig7a}
		\end{minipage}
	}
	\subfigure[]
	{
		\begin{minipage}[b]{.45\linewidth}
			\centering
			\includegraphics[height=5cm]{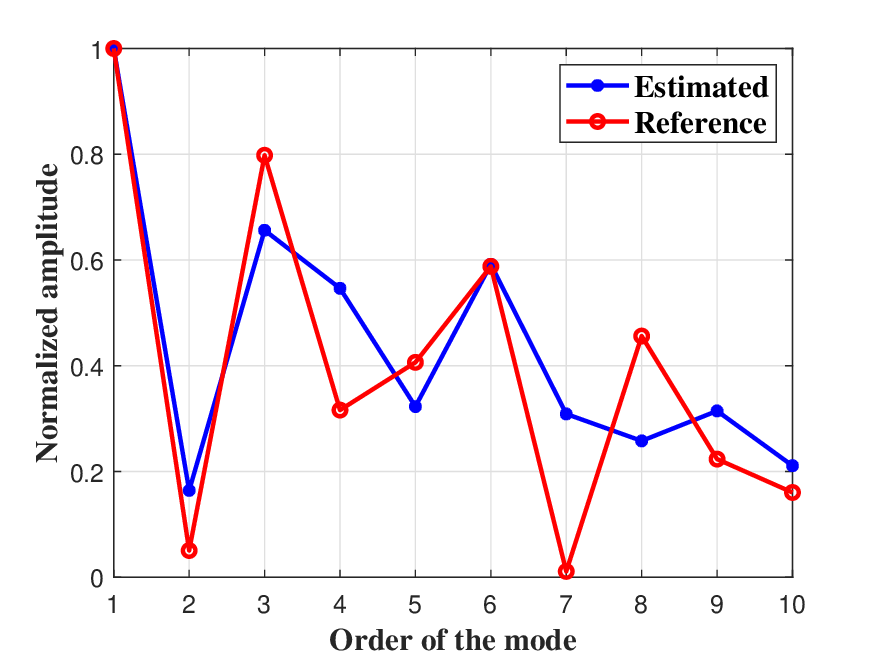}
			\vskip -5pt
			\label{fig7b}
		\end{minipage}}
        \subfigure[] 
	{
		\begin{minipage}[b]{.45\linewidth} 
			\centering
			\includegraphics[height=5cm]{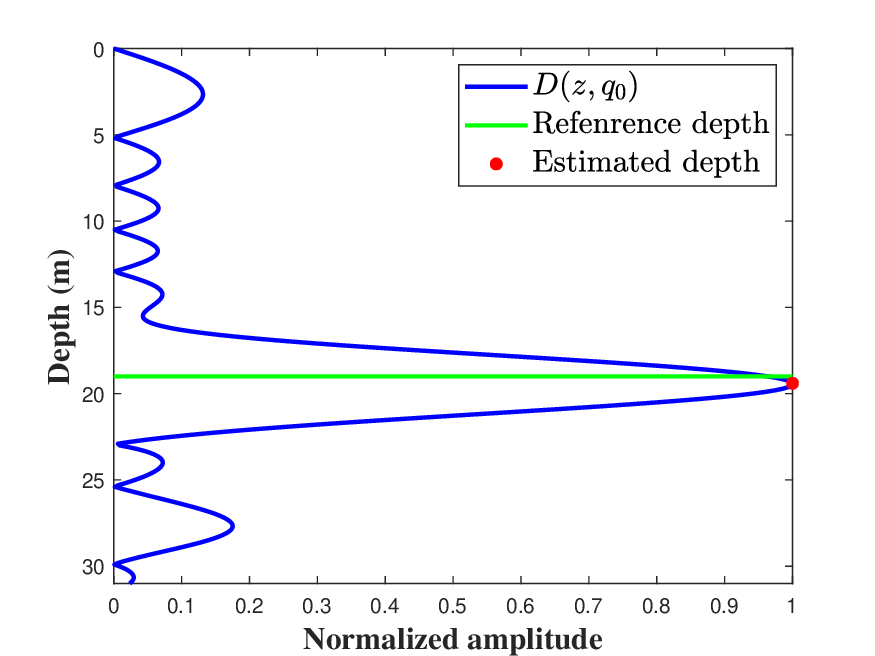}
			\vskip -5pt
			\label{fig7c}
		\end{minipage}
	}
        \subfigure[]
	{
		\begin{minipage}[b]{.45\linewidth}
			\centering
			\includegraphics[height=5cm]{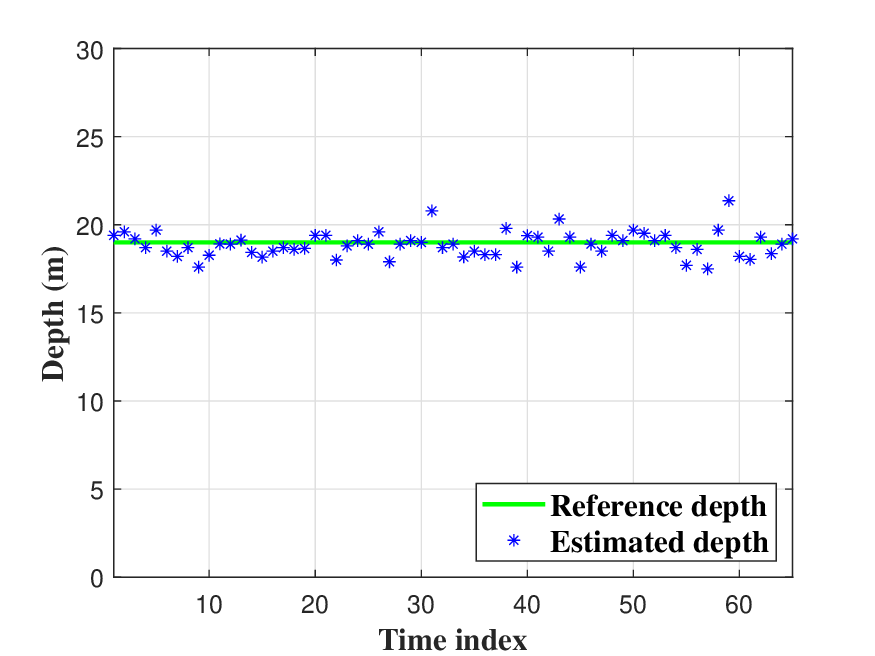}
			\vskip -5pt
			\label{fig7d}
		\end{minipage}}
	\caption{(a) Estimated normal mode wavenumbers and (b) normalized modal amplitudes at 596 Hz for a source range of 8.26 km. (c) Depth ambiguity function using OCMS-D (blue line) for a source ranged approximately 8.26 km from the VLA, estimated depth marked by a red dot. (d) The depth estimation results (blue scatter points) from processing the first 43 minutes of data received by the VLA. The reference depth is indicated by the green line. }
	\label{fig7}
\end{figure}

As an example of the Yellow Sea experiment, the data received by the VLA in the first 43 minutes—specifically, 65 segments of 4-second CW signals—were processed in this subsection, with results are shown in Fig.~\ref{fig7}. During the time window of $T = 4$ s, the source moved about 8 m, which can be considered constant relative to the distance between the source and the VLA. Consistent with Section~\ref{sec2}, the $\xi_{\min} = 2.0851$~$\rm{m^{-1}}$, $\xi_{\max} = 2.5217$~$\rm{m^{-1}}$ and the sign sampling interval in depth direction was fixed at 0.1  m. Fig. \ref{fig7a} and Fig. \ref{fig7b} present the estimated normal mode wavenumbers and normalized mode amplitudes at 596 Hz for a source range of 8.26 km. For comparison, the corresponding reference values computed from the simulation environment described in subsection \ref{subsec3.1} are also included in the figures. The high agreement observed between the estimated and reference results ensures the reliability of the subsequent source depth estimation. Fig. \ref{fig7c} illustrates the corresponding source depth estimation result. The blue curve indicates the depth ambiguity function after modal signs compensation, whereas the green line denotes the reference depth. The estimated source depth, marked by a red dot, is 19.4 m, resulting in an AE of 0.4 m. From Fig.~\ref{fig7b}  and  Fig.~\ref{fig7c}, it can be observed that although the estimated modal amplitudes exhibit errors, thus exerting limited influence on the source depth estimation. Furthermore, Fig.~\ref{fig7d} compares the depth estimation results of all 65 segments with the reference depth, where the green line represents the reference depth and blue scatter points represents the estimated depths. The estimated depths range from 17.6 m to 21.4 m, yielding a maximum absolute error of 2.4 m relative to the reference depth.

\begin{figure}[t]
	\centering
	\setlength{\abovecaptionskip}{0.cm}
	\includegraphics[height=6cm]{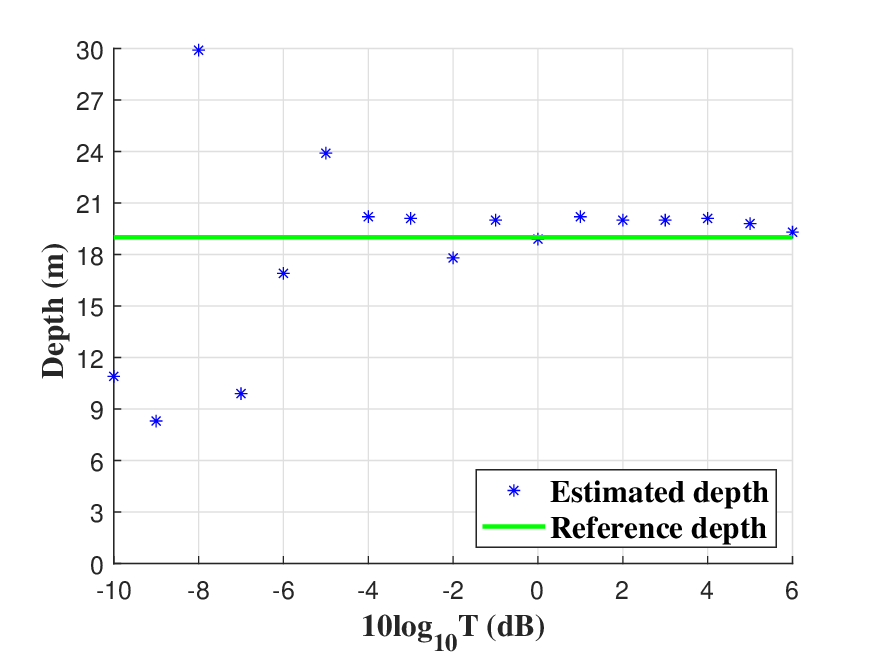}
	\caption{ Results of estimated depths under various time window lengths T. The green line represent the reference depth.}\label{fig7_2}
\end{figure}

For tonal signals, increasing the length $T$ of the time window used for Fourier transformation can improve the SNR. To more comprehensively validate the performance of OCMS-D on Yellow Sea experiment data, different $T$ were applied to process the data at a source distance of 8.26 km. And the $10\text{log}_{10}T$ varied from 6 dB down to $-10$ dB. The results of OCMS-D are shown in Fig.~\ref{fig7_2}. It can be observed that when $10\text{log}_{10}T$ exceeds -4 dB (with a time window length $T = 0.40$ s), OCMS-D yields favorable depth estimation results.

\subsection{SWellEx-96 experiment}
\label{subsec4.2}

\begin{figure}[htbp]
	\centering  
	\setlength{\abovecaptionskip}{0.cm}
	\subfigure[] 
	{
		\begin{minipage}[b]{.35\linewidth}
			\centering
			\includegraphics[height=6.5cm]{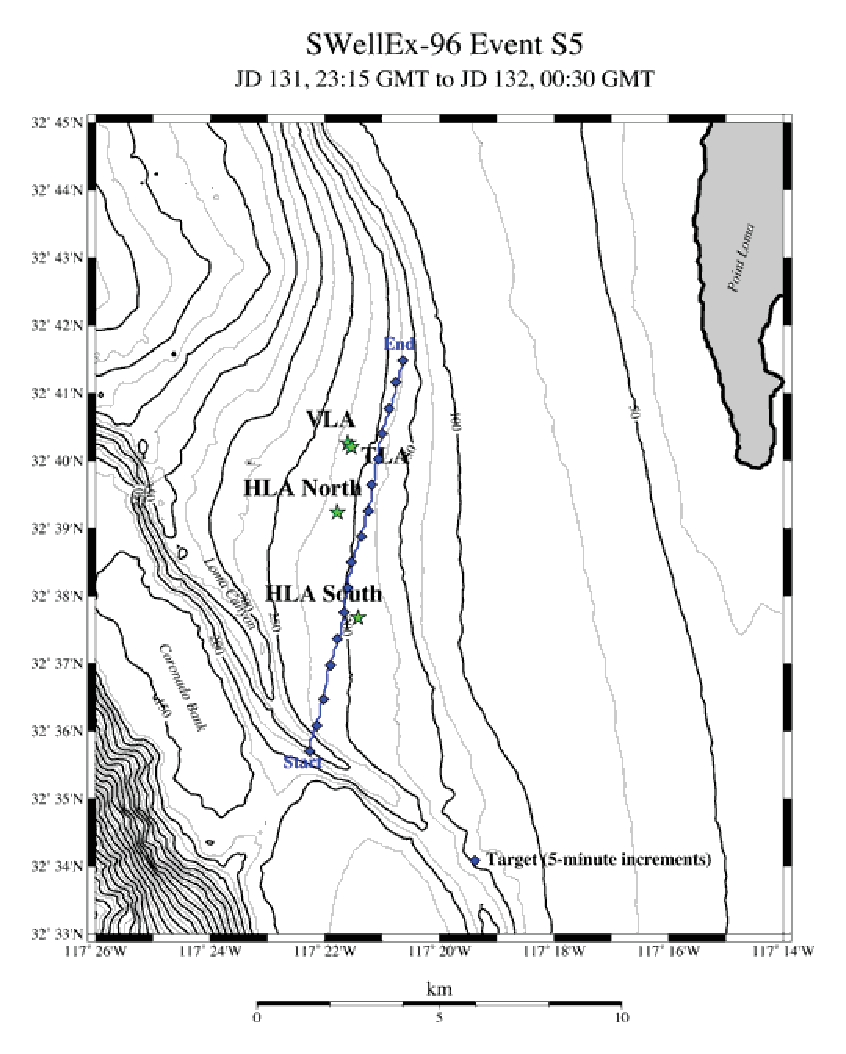}
			\vskip -5pt
			\label{fig8a}
		\end{minipage}
	}
	\subfigure[]
	{
		\begin{minipage}[b]{.55\linewidth}
			\centering
			\includegraphics[height=6cm]{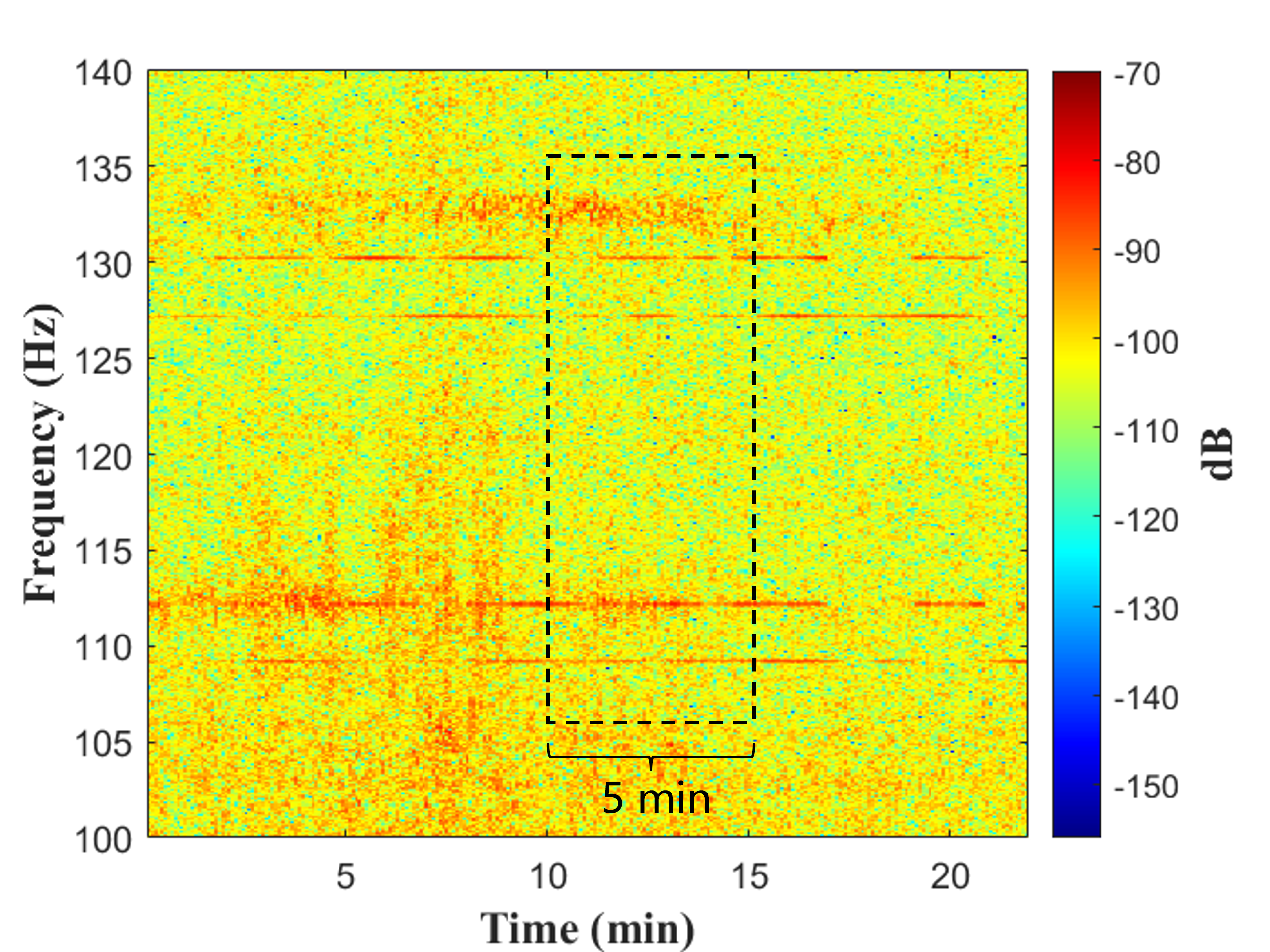}
			\vskip -5pt
			\label{fig8b}
		\end{minipage}
	}
	\caption{(a) Trajectory of the source movement for the S5 event and	the locations of hydrophone arrays. (b) Time-frequency diagram of the first 22 minutes of data received by the hydrophone closest to the water surface. And the data segment processed in this subsection is highlighted within the black box.}
	\label{fig8}
\end{figure}

Further validation of OCMS-D will be performed using S5 event data from the SWellEx-96 experiment. During this event, the source ship towed a deep source (at a depth of 54 m underwater) and a shallow source (at a depth of 9 m underwater) from southwest to northeast at 5 knots. The source trajectory is shown in Fig.~\ref{fig8a}, with most of the water depths on the trajectory ranging from 180 m to 220 m. The deep source emitted a series of tones between 49 Hz and 400 Hz, and the shallow source emitted 9 frequencies between 109 Hz and 385 Hz. In addition, a 64-array VLA (data for 21 receivers of the VLA were published on the SWellEx-96 website) was deployed during the experiment at the location shown in Fig.~\ref{fig8a}, at a water depth of 216.5m, spanning the bottom half of the water column. The lower-frequency signals from both deep source and shallow source were processed. Specifically, we choose 112 Hz and 130 Hz frequencies for the deep source, and 109 Hz and 127 Hz frequencies for the shallow source in data processing. To obtain more clearer line spectra data for all four frequencies, a data segment spans 5 minutes are processed as an example of the SWellEx-96 experiment, specifically from 10 minutes after the experiment start to 15 minutes after the experiment start, as indicated by the black box in Fig.~\ref{fig8b}. During processing, the frequency-domain pressure was calculated by the Fourier transform with $T = 10$ s, resulting in 31 data segments. Similar to the Yellow Sea experiment, the distance of the sources motion is small (about 25 m) and can be considered constant relative to the distance between the source and the VLA.

\begin{figure}[htbp]
	\centering  
	\setlength{\abovecaptionskip}{0.cm}
	\subfigure[] 
	{
		\begin{minipage}[b]{.45\linewidth} 
			\centering
			\includegraphics[height=5cm]{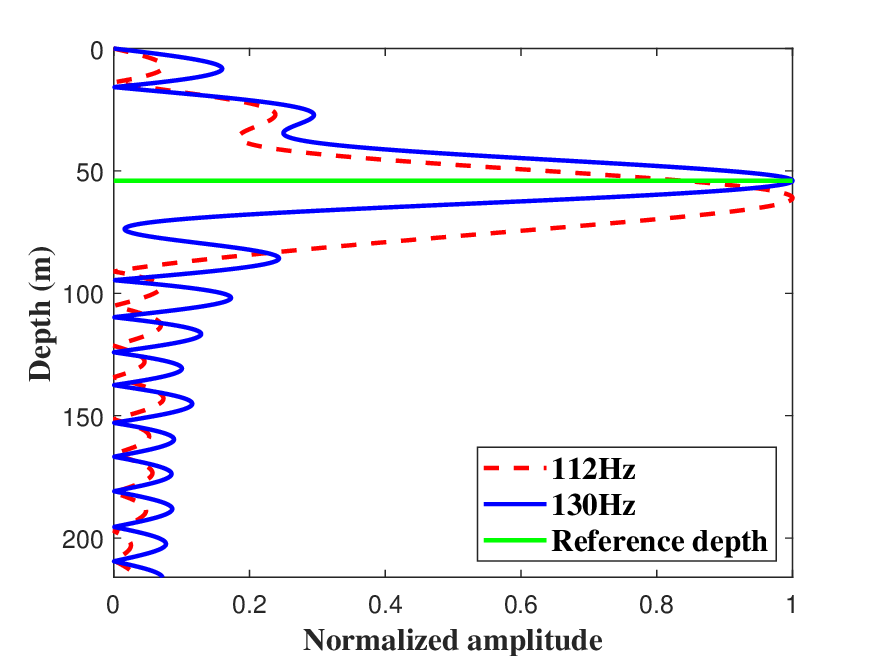}
			\vskip -5pt
			\label{fig9a}
		\end{minipage}}
	\subfigure[]
	{
		\begin{minipage}[b]{.45\linewidth}
			\centering
			\includegraphics[height=5cm]{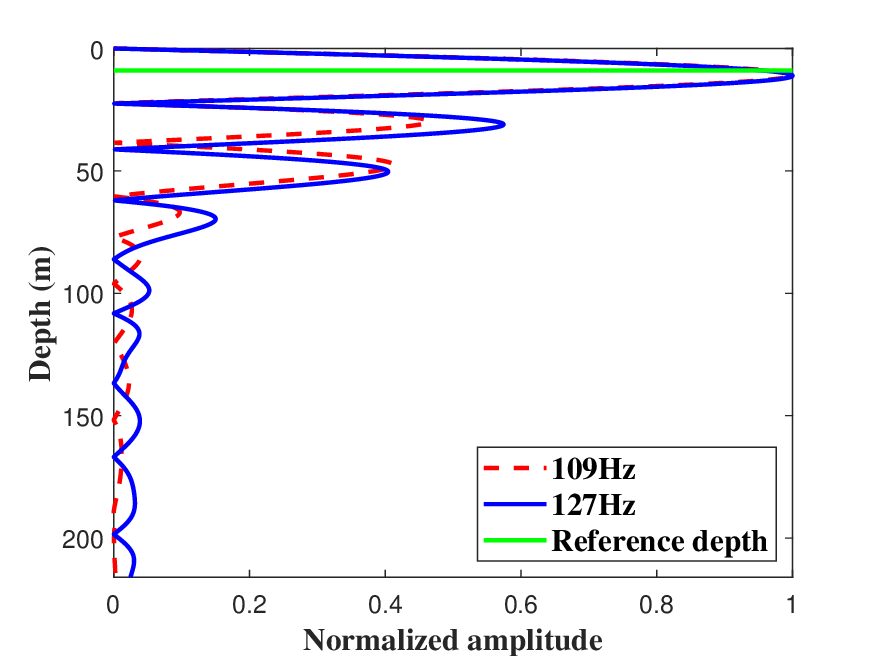}
			\vskip -5pt
			\label{fig9b}
		\end{minipage}}
	\caption{The depth ambiguity functions at the four frequencies estimated by the OCMS-D when the sources ranged approximately 6.89 km from the VLA. (a) The deep source’s depth ambiguity functions at 112 Hz (red dashed line) and 130 Hz (blue line). (b) The shallow source’s depth ambiguity functions at 109 Hz (red dashed line) and 127 Hz (blue line). The green line indicates the reference depth.}
	\label{fig9}
\end{figure}

\begin{figure}[htbp]
	\centering  
	\setlength{\abovecaptionskip}{0.cm}
	\subfigure[] 
	{
		\begin{minipage}[b]{.45\linewidth} 
			\centering
			\includegraphics[height=5cm]{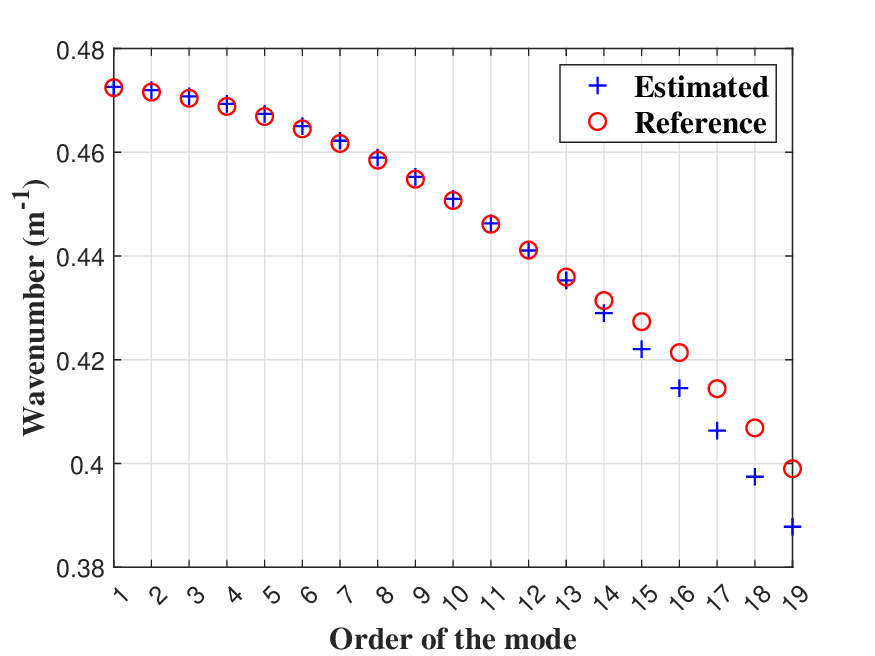}
			\vskip -5pt
			\label{fig9_2a}
		\end{minipage}}
	\subfigure[]
	{
		\begin{minipage}[b]{.45\linewidth}
			\centering
			\includegraphics[height=5cm]{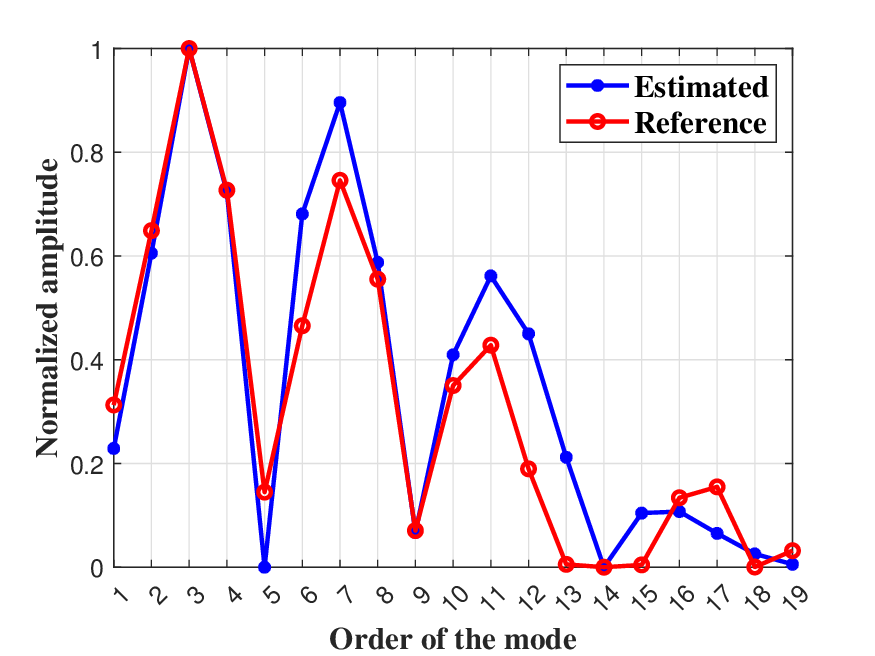}
			\vskip -5pt
			\label{fig9_2b}
		\end{minipage}}
        	\subfigure[] 
	{
		\begin{minipage}[b]{.45\linewidth} 
			\centering
			\includegraphics[height=5cm]{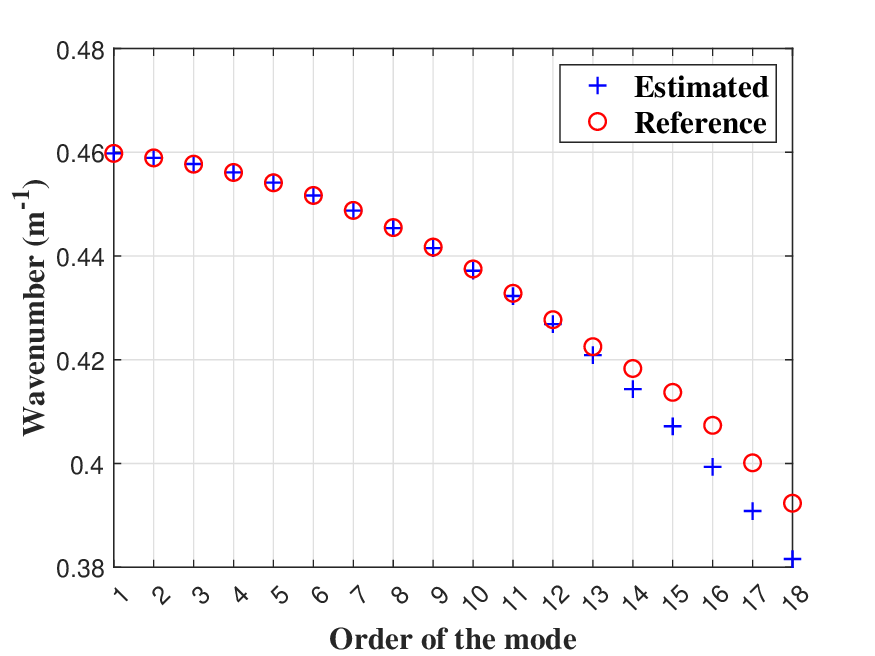}
			\vskip -5pt
			\label{fig9_2c}
		\end{minipage}}
	\subfigure[]
	{
		\begin{minipage}[b]{.45\linewidth}
			\centering
			\includegraphics[height=5cm]{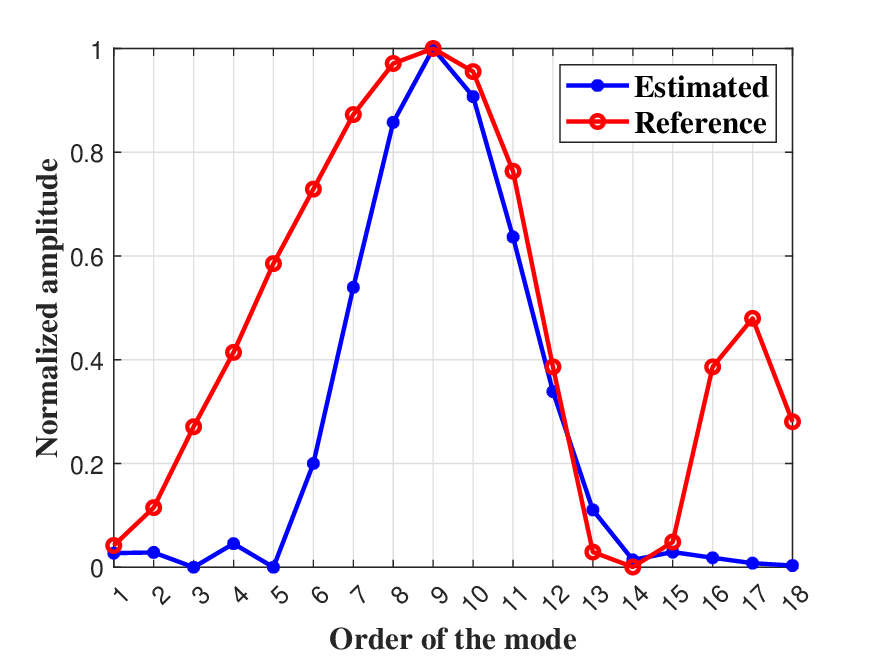}
			\vskip -5pt
			\label{fig9_2d}
		\end{minipage}}
	\caption{Estimated (a) normal mode wavenumbers and (b) normalized modal amplitudes at 112 Hz for the deep source range of 6.89 km. Estimated (c) normal mode wavenumbers and (d) normalized modal amplitudes at 109 Hz for the shallow source range of 6.89 km.}
	\label{fig9_2}
\end{figure}

In this subsection, we choose $\xi_{\min} = 2{\pi}f/1822$~$\rm{m^{-1}}$, $\xi_{\max} =2{\pi}f/1488$~$\rm{m^{-1}}$ for different frequencies. And the sign sampling interval in depth direction was fixed at 0.1 m. Fig.~\ref{fig9} illustrated the depth ambiguity functions for the deep source and shallow source estimated by the OCMS-D when the sources ranged approximately 6.89 km from the VLA. Fig.~\ref{fig9a} indicated that deep source’s depth ambiguity functions at 112 Hz (red dashed line) and 130 Hz (blue line), with the estimated source depths are 53.5 m and 61.1 m, respectively. Similarly, Fig.~\ref{fig9b} indicated that shallow source’s depth ambiguity functions at 109 Hz (red dashed line) and 127 Hz (blue line), with the estimated source depths are 10.9 m and 11.1 m, respectively. The source depth estimation results in Fig.~\ref{fig9} are further analyzed below, taking the deep source at 112 Hz and the shallow source at 109 Hz as examples. Fig.~\ref{fig9_2} (a) and (b) present the estimated normal‑mode wavenumbers and normalized modal amplitudes at 112 Hz for the deep source at a range of 6.89 km, while Fig.~\ref{fig9_2} (c) and 12(d) show the corresponding estimates for the shallow source at 109 Hz and the same range. For comparison, reference results simulated using the SWellEX‑96 environmental parameters given in ~\cite{ref25} are also provided in Fig.~\ref{fig9_2} . It can be observed that although the wavenumber estimates of higher‑order modes (mode numbers greater than 12) exhibit relatively large errors, their corresponding amplitudes are low, thus exerting limited influence on the source depth estimation. Furthermore, compared with the deep source, the shallow source excites normal modes that are predominantly concentrated in higher‑order modes, which is consistent with the normal mode theory. 


\begin{figure}[t]
	\centering
	\setlength{\abovecaptionskip}{0.cm}
	\includegraphics[height=6cm]{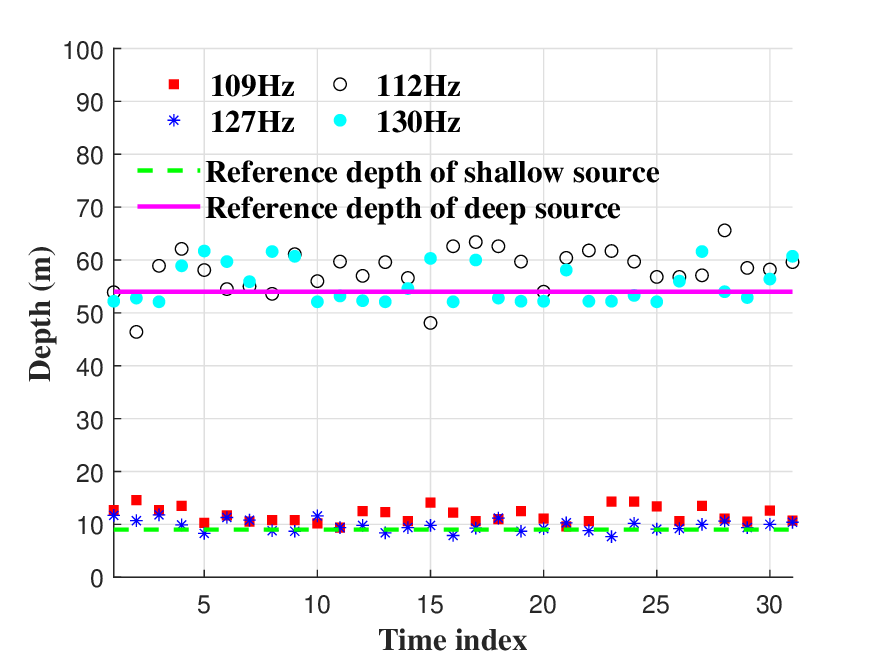}
	\caption{ Results of estimated depths for the shallow sources at 109 Hz (marked by red dots) and 127 Hz (marked by blue dots) and for the deep sources at 112 Hz (marked by black dots) and 130 Hz (marked by cyan dots). The green line and magenta line represent the reference depth of the shallow source and deep source, respectively.}\label{fig10}
\end{figure}

Furthermore, the results of all 31 data segments for the deep source and shallow source are shown in Fig.~\ref{fig10}. In this figure, red squares and black circles represent the measurement results for the lower frequencies (109 Hz and 112 Hz), while blue asterisks and cyan scatter points denote the results for the higher frequencies (127 Hz and 130 Hz). The green dashed line and the magenta solid line indicate the reference depths for the shallow source and the deep source, respectively. The results for the two frequencies of the deep source both fall within the range of 46.4 m to 64.8 m, with a maximum AE is 10.8 m. And the results for the two shallow-source frequencies range from 7.7 m to 14.4 m, with a maximum AE is 5.4 m. Moreover, the results at 127 Hz generally outperform those at 109 Hz, likely due to the 127 Hz frequency being able to excite more normal modes. Overall, OCMS-D also demonstrates good performance in experimental data.

\begin{figure}[htbp]
	\centering  
	\setlength{\abovecaptionskip}{0.cm}
	\subfigure[] 
	{
		\begin{minipage}[b]{.45\linewidth} 
			\centering
			\includegraphics[height=5cm]{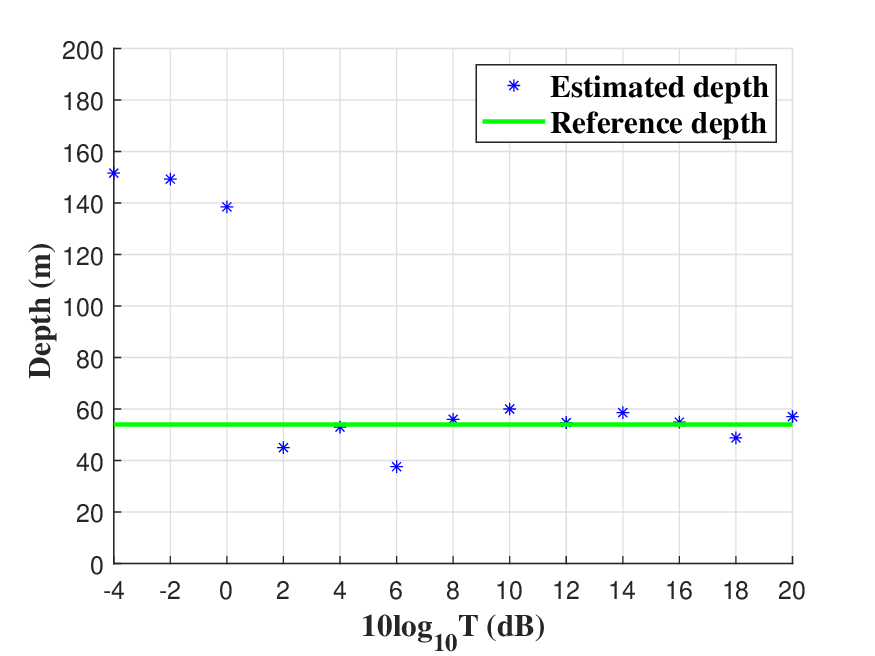}
			\vskip -5pt
			\label{fig10_2e}
		\end{minipage}}
	\subfigure[]
	{
		\begin{minipage}[b]{.45\linewidth}
			\centering
			\includegraphics[height=5cm]{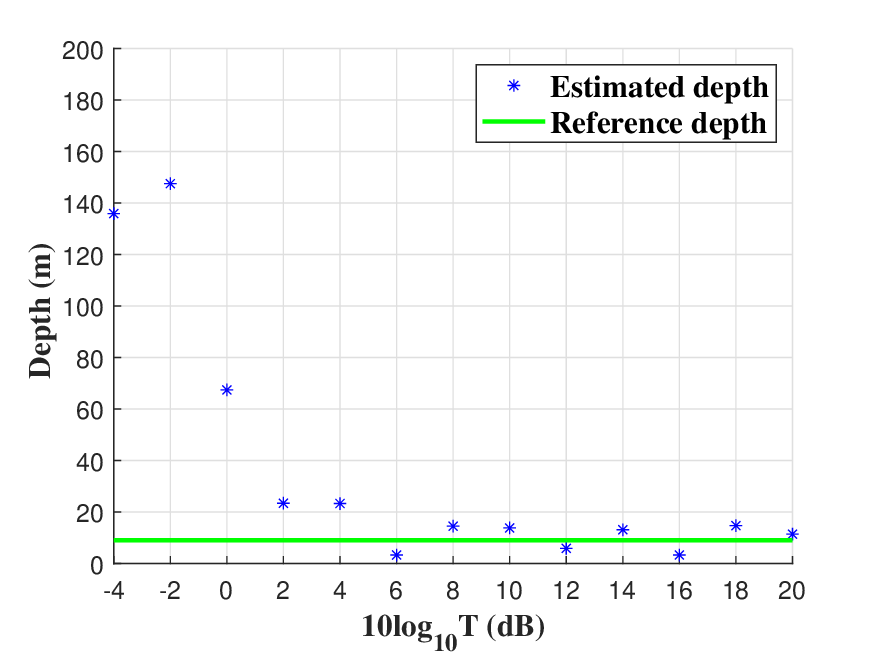}
			\vskip -5pt
			\label{fig10_2f}
		\end{minipage}}
	\caption{Results of estimated deep and shallow source depths under various time window lengths T. The green line represent the reference depth.}
	\label{fig10_2ef}
\end{figure}

Finally, using data from deep and shallow sources at lower frequencies (112 Hz and 109 Hz) when the source-VLA distance about 6.89 km as an example, the SWellEX-96 experimental data were similarly processed with different time window lengths $T$.  The $10\text{log}_{10}T$ varied from 20 dB down to $-4$ dB. The results are shown in Fig.~\ref{fig10_2ef}. It can be observed that when $10\text{log}_{10}T$ exceeds 2 dB (with a time window length $T = 1.58$s), OCMS-D yields favorable depth estimation results.

\section{Conclusion}
\label{sec5}

This paper proposes the OCMS-D method, which enables depth estimation of a fixed-range source using a VLA with single-frequency signals in an environment with unknown seabed properties, without requiring the array to span the entire water column. The performance and effectiveness of OCMS-D are evaluated through numerical simulations and sea-trial data. Results indicate that, in general, the source depth estimation accuracy of OCMS-D improves with higher SNR, larger VLA aperture, and more array elements. However, simulation studies reveal that when the VLA aperture is small, the estimation accuracy depends on the array deployment location. Furthermore, when the VLA position is fixed, the accuracy does not decrease monotonically with a reduction in the number of array elements. This suggests that for a given aperture, an optimal VLA design and deployment strategy may exist specifically for OCMS-D, although such an investigation falls beyond the scope of this paper and will be addressed in future work.

\section*{CRediT authorship contribution statement}
\textbf{Yangjin Xu:} Writing - original draft, Methodology, Investigation, Conceptualization, Writing - review \& editing.
\textbf{Wei Gao:}  Writing - review \& editing,  Supervision, Funding acquistion,  Formal analysis, Validation. 
\textbf{Xiaolei Li:} Writing - review \& editing, Methodology, Conceptualization, Supervision, Funding acquistion,  Formal analysis.
\textbf{Qinghang Zeng:} Supervision, Investigation.

\section*{Data availability statement}
Data will be made available on request.

\section*{Declaration of Competing Interest}
The authors declare that they have no known competing financial interests or personal relationships that could have appeared to influence the work reported in this paper.
    
\section*{Acknowlgedgments}
This work is supported by the National Natural Science Foundation of China (Grant No. 12474444), and the Open Research Fund of Hanjiang National Laboratory (Grant No. KF2024041).

\bibliography{reference}

\end{document}